\documentclass{aa}
\usepackage{natbib}
\usepackage{graphicx}
\usepackage{epsfig}
\usepackage{amsmath,amssymb}
\usepackage[dvipdfm,breaklinks=true,bookmarks=true,hypertexnames=false,bookmarksnumbered=true,bookmarksopen=false,bookmarks=true,bookmarksnumbered=true]{hyperref} 
\usepackage[dvips]{thumbpdf}

\bibpunct{(}{)}{;}{a}{}{,}

\newcommand{\coordra}[3]{#1^{\rm h}~#2^{\rm m}~#3^{\rm s}}
\newcommand{\coorddec}[3]{#1^{\rm o}~#2{\rm '}~#3{\rm ''}}
\newcommand{\reffig}[1]{Fig.~\ref{#1}}
\newcommand{\refsec}[1]{Sect.~\ref{#1}}
\newcommand{\hm}{\,h^{-1}_{70}}
\newcommand{\hmMpc}{\hm\,{\rm Mpc}}
\newcommand{\hmkpc}{\hm\,{\rm kpc}}
\newcommand{\hmMsun}{\hm\,{\rm M}_\odot}
\newcommand{\mslsun}{\,h_{70}\,({\rm M/L})_\odot}
\newcommand{\der}{\,{\rm d}}
\newcommand{\eexp}[1]{\,{\rm e}^{#1}}
\newcommand{\mypm}[2]{^{+#1}_{-#2}}
\newcommand{\mycov}[2]{ \langle \kappa_{#1}\kappa_{#2} \rangle }
\newcommand{\mycovv}[1]{ \langle \kappa_{#1}^2 \rangle }

\newcommand{\scrit}{\Sigma_{\rm crit}}
\newcommand{\vs}{{\sl vs.~}}
\newcommand{\eg}{{\sl e.g.}}

\begin{document}
\def\sizex{16.0 cm}
\def\bigx{10.0 cm}
\def\smallerxsize{7.0 cm}
\def\smallxsize{10.0 cm}
\def\smallysize{12.0 cm}

\title{Mass and Light in the Supercluster of Galaxies MS0302+17
  \thanks{Based on observations obtained at the Canada-France-Hawaii
    Telescope (CFHT) which is operated by the National Research
    Council of Canada, the Institut des Sciences de l'Univers of the
    Centre National de la Recherche Scientifique and the University of Hawaii.}}
\titlerunning{Mass and Light in the MS0302+17 Supercluster}
\authorrunning{Gavazzi et al.}

\author{R. Gavazzi \inst{1} \and Y. Mellier \inst{1,2} \and  B. Fort\inst{1}
  \and J.-C. Cuillandre \inst{3} \and M. Dantel-Fort \inst{2}}

\offprints{R. Gavazzi, \email{gavazzi@iap.fr}}
\date{}
\institute{Institut d'Astrophysique de Paris, UMR 7095, 98 bis Boulevard Arago,
  75014 Paris, France \and Observatoire de Paris, LERMA, UMR 8112, 61 Avenue de
  l'Observatoire, 75014 Paris, France \and Canada-France-Hawaii Telescope,
  65-1238 Mamalahoa Highway, Kamuela, HI 96743 }

\abstract{We investigate the supercluster MS0302+17 ($z \approx 0.42$)
  using weak lensing analysis and deep wide field $BVR$ photometry with the
  CFH12K camera. Using $(B-V)$ \vs $(V-R)$ evolution tracks we identify
  early-type members of the supercluster, and foreground ellipticals.
  We derive a R band catalogue of background galaxies for weak lensing analysis.
  We compute the
  correlation functions of light and mass and their cross-correlation and
  test if light traces mass on supercluster, cluster and galaxy scales.\\
  We show that the data are consistent with this assertion. The $\zeta$-statistic
  applied in regions close to cluster centers and global correlation
  analyses over the whole field converge toward the simple relation
  $M/L=300\pm30\mslsun$ in the B band. This independently confirms 
  the earlier results obtained by \citet{Kaiser98}.\\
  If we model dark matter halos around each early-type galaxy by a truncated
  isothermal sphere, we find that a linear relation $M\propto L$ still holds. 
  In this case, the average halo truncation radius is $s_* \lesssim 200\hmkpc$
  close to clusters cores whereas it reaches a lower limit of $\sim 300\hmkpc$
  at the periphery. This change of $s_*$ as a function of radial distance
  may be interpreted as a result of tidal stripping of early type galaxies.
  Nevertheless the lack of information on the spatial distribution of
  late-type galaxies affects such conclusions concerning variations of $s_*$.\\
  Though all the data at hand are clearly consistent with the assumption that mass
  is faithfully traced by light from early-type galaxies, we are not able
  to describe in detail the contribution of late type galaxies.
  We however found it to be small. Forthcoming wide surveys in UV,
  visible, and near infrared wavelengths will provide large enough samples to extend
  this analysis to late-type galaxies using photometric redshifts.
}
\maketitle

\section{Introduction}\label{sec:introduction}
Detailed investigations of superclusters of galaxies help to understand
the late evolution of large scale structures in the transition phase
between the linear and non-linear regime.
In contrast to wide field cosmological surveys that primarily draw
the global structure formation scenario, supercluster studies focus
on more detailed descriptions of the physical properties of baryonic and
non-baryonic matter components on tens of kiloparsec to tens of megaparsec scales.
Within the evolving cosmic web, numerical simulations predict that gas cooling 
or dark halos and galactic interaction processes start prevailing against
large scale gravitational clustering and global expansion of the universe
\citep{vogeley94,bond96,kauffmann98}. Small scale dissipation processes 
transform early relations between the apparent properties of 
large scale structures and those of the underlying matter
content. The comparison of mass-to-light ratios and of the mass 
and light distributions as a function of local environment between 
supercluster and cluster scales may therefore reveal useful imprints
on the physical processes involved in the generation of linear and non-linear
biasing \citep{kaiser84,bardeen86}.

The properties of large scale structures (LSS)  
can be characterized by  optical, X-ray, Sunyaev-Zeldovitch effect 
and weak lensing observations. These techniques are widely 
used on clusters and groups of galaxies 
\citep[see \eg][]{hoekstra01,carlberg,bahcall95}, but 
their use is still marginal
at larger scales. Superclusters of galaxies are therefore still 
poorly known systems. Besides early investigations  
\citep{davis80,postman88,quintana95,small98}, all recent
studies on Abell901/902 \citep{Gray02} (hereafter G02), Abell222/223
\citep{proust00,dietrich02}, MS0302+17
\citep{fabricant,Kaiser98} (hereafter K98),
Cl1604+4304/Cl1604+4321 \citep{lubin00} or
RXJ0848.9+4452 \citep{rosati} show that 
superclusters of galaxies are genuine physical systems 
where gravitational interactions between clusters of galaxies
prevail. There is however no conclusive evidence that 
superclusters are gravitationally bound systems.

Numerical simulations show that supercluster properties may be best
characterised by the shape and matter content of the filamentary structures
connecting neighboring groups and clusters. Unfortunately, the physical
properties of these filaments are still poorly known, though their existence 
at both low and very high redshift seems confirmed by 
a few optical and X-ray observations \citep{moller01,durret03}.

An alternative approach has been proposed by K98,
G02 and \citet{clowe98} who used weak lensing analyses.
While gravitational lensing has been used on a large sample of clusters of
galaxies \citep[See \eg][and references therein]{Mellier99,BartShneid01},
its use at supercluster scales is still in its infancy.
It has been pioneered by K98 and \citet{Kaiser99} who used V and I band data
obtained at CFHT  to probe the matter distribution in the MS0302+17
supercluster. Similar analyses were done later by G02 for the A901/902 system
using wide field images obtained with the WiFi instrument mounted at the
MPG/ESO 2.2 telescope in La Silla Observatory. \citet{wilson01} applied similar
techniques to ``empty'' fields. In these papers, the projected mass density,
as reconstructed from the distortion field of background galaxies,
has been compared to the light distribution on several scales.
All these studies conclude that there is a strong relation 
between the light of early type galaxies the and dark matter distributions
as if all the mass was traced by early type galaxies.

Over scales of a few megaparsecs, K98 found that the dark matter does not extend
further than the light emission derived from the early-type galaxies sample.
According to their study, there is almost no mass associated with late type
galaxies. In contrast, G02 argued that there is some,
but its fraction varies from one cluster to another,
leading to a lower and more scattered $M/L$ than K98 found.
This discrepancy may be explained if either the two superclusters
are at different dynamical stages, or their galaxy populations differ
(fraction of early/late type galaxies).

The reliability of their results may however strongly depend on systematic
residuals. It is worth noticing that the weak lensing signal produced
by filaments is indeed expected to be difficult to detect because projection
effects may seriously dilute the lensing signal \citep{jain02}.
Therefore, systematics produced by technical problems related to the way 
both groups analyzed their data may also be a strong limitation. 
It is therefore important to confirm early K98 conclusions from an independent
analysis, and possibly go further in order to take into account 
the properties of dark halos. In particular, it is interesting to 
compare the halo properties (size, velocity dispersion) of cluster galaxies
with those of field galaxies. MS0302+17 seems a generic and almost ideal
$z\approx0.42$
supercluster configuration for such an astrophysical study because it is composed
of three very close rich clusters (mean projected distance $\sim15\arcmin$).

In this paper, we describe the investigation of this system.
Using new data sets obtained in B, V and R at the CFHT with the CFH12K CCD camera,
we explore the mass and light distributions in the supercluster area.
Since the CFH12K field of view is 1.7 times larger than the UH8k camera,
the three clusters and most of their periphery  are totally encompassed in
the CFH12K field and we can even explore whether other clusters lie in the
field at the same redshift. A quick visual inspection of the southern cluster
(ClS) reveals that it is very dense. The giant arc discovered by \citet{Mathez92}
is visible. Similar arc(let) features are also visible in the northern
(ClN) and eastern (ClE) systems, making the MS0302+17 supercluster a unique
spectacular lensing configuration, where strong and weak lensing inversions
can be done.

This paper is organized as follows. In \refsec{sec:thedata} we
describe the observations that were carried out and reduced. We include
details on how both astrometric and photometric solutions were computed.
We present a detailed quality assessment of the catalogues,
where comparisons to existing deep catalogues are made.
In \refsec{sec:cluster-photom} we explain how object catalogues were
produced for supercluster members identification. \refsec{sec:weak-lensing}
presents the weak lensing signal produced by the dark matter component.
\refsec{sec:correlation} shows how these two components cross-correlate.
Results are discussed in \refsec{sec:summary} and our conclusions and
summary are presented in \refsec{sec:conclusion}.
Throughout this paper we adopt the cosmology
$\Omega_m=0.3,\quad \Omega_\Lambda=0.7,
\quad H_0 = 70\,h_{70}\,{\rm km.s}^{-1} {\rm Mpc}^{-1}$,
leading to the scaling relation $1 \arcmin = 333 \hmkpc$ at $z=0.42$.

\section{The Data}\label{sec:thedata}
\subsection{Observations and data reduction}\label{subsec:obsdatared}
The observations of the MS0302+17 supercluster area were obtained on
October 12, 1999.  They were carried out with the CFH12K camera mounted
at the prime focus of the Canada-France-Hawaii Telescope. The CFH12K mosaic device
is composed of $6 \times 2 $ thinned backside illuminated MIT Lincoln
Laboratories CCID20 $2048\times 4096$ CCDs with a $15\,\mu\rm m$ pixel size. 
The wide field corrector installed at the prime focus provides an
average pixel scale of $0\farcs205$ and the whole field of view of the CFH12K
camera is $43\farcm2\times 28\farcm9$.
Useful details on the camera can be found in
\citet{cuillandre} and in \citet{Mccracken03} (hereafter McC03).

The pointings were centered at the reference position
RA(2000) = $\coordra{03}{05}{26.00}$ and DEC(2000) = $\coorddec{+17}{17}{54}$,
so that the CFH12K field of view encompasses the
three major clusters of galaxies.
Sequences of dithered exposures were obtained in B, V and R filters
using small shifts of about 30 arcsec to fill the gaps
between the CCDs and to accurately flat field each individual
image frame. Table \ref{tab:obs} summarizes and assesses the
useful observations used in this paper.

The B, V and R data were processed and calibrated at the TERAPIX data center
located at IAP. The pre-calibration process was done using the {\tt flips}
package\footnote{\url{http://www.cfht.hawaii.edu/~jcc/Flips/flips.html}}.
Photometric and astrometric calibrations, as well as image stacking
and catalogue production were done using the current software package
available at TERAPIX\footnote{\url{http://terapix.iap.fr/}}.
The overall pre-reduction (bias and dark subtraction, flat-field calibration),
photometric and astrometric calibration as well as image resampling and
co-addition follow exactly the same algorithms and steps as in
McC03. We refer to this paper for further details.

\begin{table}
  \centering
  \caption{Summary of the observations with total exposure time
    (+ the number of dithered pointings), seeing and limiting
    magnitude. Following \citet{Mccracken03}, the $AB$ limiting magnitude
    corresponds to a 50\% completeness limit.}
  \begin{tabular}{*{8}{c}}
    Filter & exp. time & Seeing   & Limiting   \\
    & (seconds) & (arcsec) &   Mag      \\\hline
    $B_{AB}$ & $6 \times 600 = 3600$ & 0.9 & 25.75 \\
    $V_{AB}$ & $6 \times 600 = 3600$ & 0.9 & 25.50 \\
    $R_{AB}$ & $10\times 720 = 7200$ & 0.7 & 26.50 \\
  \end{tabular}
  \label{tab:obs}
\end{table}

\subsection{Photometric calibration}\label{subsec:photometry}
The photometric calibration was done using Landolt star fields SA95 and SA113
\citep{Landolt92}. The IRAS maps \citep{Schlegel98} show that the Galactic
extinction is important in this field. The average $E(B-V)=0.125$, leading to
extinction corrections in $B$, $V$ and $R$ of 0.508, 0.384 and 0.285 respectively.
We adopted the $AB$ magnitude corrections provided by McC03:
$B_{AB}=B-0.097$, $V_{AB}=V-0.007$ and $R_{AB}=R+0.218$.

\begin{figure}
  \resizebox{\hsize}{!}{\includegraphics{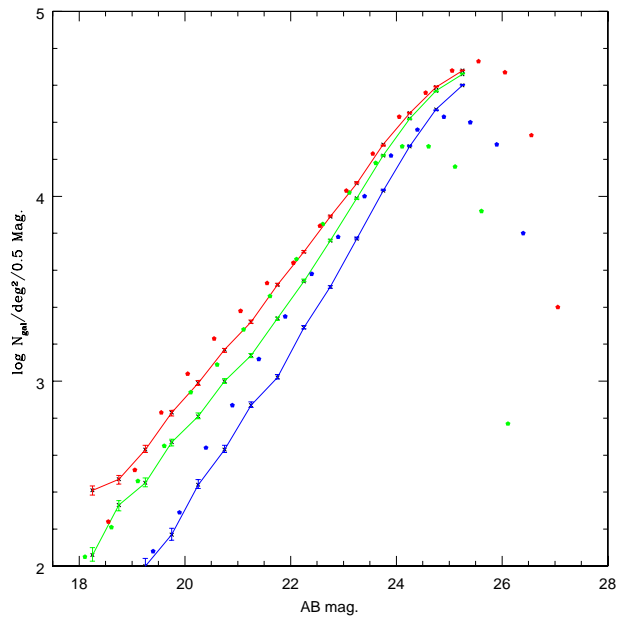}}
  \caption{$BVR$ galaxy counts in the MS0302+17 supercluster
    field. The red, green and blue dots show the galaxy number density
     in  $R$, $V$ and $B$, respectively.
    The solid red, green and blue lines show the same counts
    obtained with the same filters on the deep VIRMOS F02 field by
    McC03. For the bright and faint magnitudes the agreements
    are excellent. In between, the MS0302+17 counts show a systematic excess
    that reveals the supercluster populations at $z =0.42$.}
  \label{fig:galcountms0302}
\end{figure}

The object photometry was derived using the {\tt Magauto} parameter
of {\tt SExtrator} \citep{Bertin96} for the magnitude and {\tt Magaper}
for the color index, with the V catalogue as reference position, inside
a 2 arcsec aperture. The galaxy number counts derived from this
photometry peak at $B_{AB}=24.9$, $V_{AB}=24.3$ and $R_{AB}=25.6$.
As in McC03, we derived the limiting magnitude by adding
simulated stellar sources in the field. The limiting depths at which 50\% of
these sources are recovered are $B_{AB}=24.9$, $V_{AB}=24.3$ and $R_{AB}=25.6$,
in good agreement with the expectations from the McC03 F02 deep exposures,
once rescaled to a similar exposure time (see \reffig{fig:galcountms0302}).

The reliability of the photometric calibration was checked by
comparing the $B$, $V$ and $R$ galaxy counts to those published for
the VIRMOS F02 field by McC03. The supercluster galaxy
population contaminates the galaxy number counts. However, as the
$B-(B-V)$ and $R-(B-R)$ color-magnitude diagrams show, the bright-end
and faint-end galaxy populations are dominated by field galaxies,
so we expect the data of the MS0302+17 supercluster and the VIRMOS F02 data
to be comparable for these populations. We checked that the counts agree
within 0.05 magnitudes with McC03 for the three filters.

\begin{figure}
  \resizebox{\hsize}{!}{\includegraphics{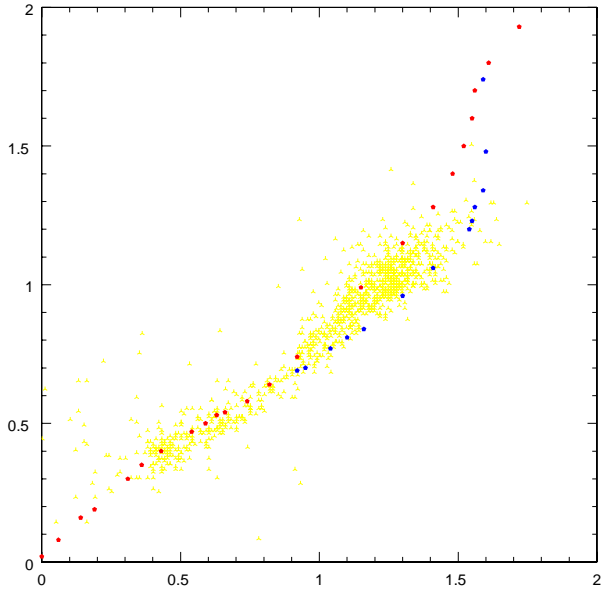}}
  \caption{Color-color diagram of bright stellar objects in the MS0302+17
    supercluster field. The horizontal and vertical axes are the
    Johnson $(B-V)$ and $(V-R)$ color indices, respectively.
    The red and blue dots show the stellar tracks for Main Sequence
    and Giant stars, as predicted by \citet{Johnson66}. The yellow dots
    are the color positions of the bright star sample.
    The figure clearly shows two populations : the blue halo stars
    ($(B-V)<1.0$) and the red disk M-dwarf stars.}
  \label{fig:stelalrcolorms0302}
\end{figure}

We also checked the $(B-V)$ versus $(V-R)$ color-color magnitude of stars,
as selected from the {\tt SExtractor} stellar index. To avoid mixing of 
galaxies in the sample, we only used bright objects that have a
reliable stellar index. The colors are plotted in
\reffig{fig:stelalrcolorms0302} (yellow dots) and compared to selected
stars on the main sequence and giant-star $(B-V)/(V-R)$ tracks
from \citet{Johnson66}. Two populations are clearly visible:
the blue stars ($B-V<1.0$) are halo stars, the red ones
($\langle B-V \rangle \simeq 1.3$) are disk M-dwarf stars.
The blue population perfectly matches the Main Sequence, but for the red sample
we found a systematic offset of $0.1$ in the $(B-V)$ term.
A similar discrepancy has already been mentioned by \citet{Prandoni99}.
It is likely due to the color correction, which is no longer linear
for those stars. The dispersion is easily explained by the intrinsic
dispersion of stellar colors and likely from statistical magnitude
measurement errors. We therefore conclude that our $(B-V)$ and $(V-R)$
colors have an internal error of $\pm 0.05$ magnitude.

The CCD to CCD calibration errors are in principle minimized since
all CCDs are rescaled with respect to the reference CCD$\#$4 that contains
several Landolt stars.
However, residuals from illumination correction may still bias the calibration.
We checked the CCD to CCD stability by comparing the B, V and R galaxy counts.
The contamination by cluster galaxies in the field
makes this approach difficult. Their effects were minimized by
removing the cluster regions from the count estimates. However,
since the diffuse supercluster filaments also contaminate the signal,
large CCD to CCD fluctuations of the counts still remain. We therefore
focused on the faint end part of the magnitude distribution, where the
supercluster populations should be negligible.
Using these constraints, the average CCD to CCD galaxy count
fluctuations in B, V and R are 2.5\% in each filter, with peaks of 7.5\%.
When clusters fields are included and the magnitude range is
broadened, the peaks reach 16\%. This clearly reveals the presence of
clusters populations. Possible residuals from calibration problems are
therefore negligible compared to fluctuations expected from Poisson or
cosmic variance.

Possible color variations of stars from one CCD to another were checked 
using a color analysis of 100 stars per chip. We found that they all show similar
average $(B-V)$ and $(V-R)$ colors inside the 12 CCDs, within $\pm 0.1$
($ie\:2\sigma$) and without any systematic color gradient. This confirms,
independently of the galaxy counts, that the photometry and our $(B-V)$ and
$(V-R)$ colors are stable enough across the field to meet our scientific goals.
\begin{figure}[bth]
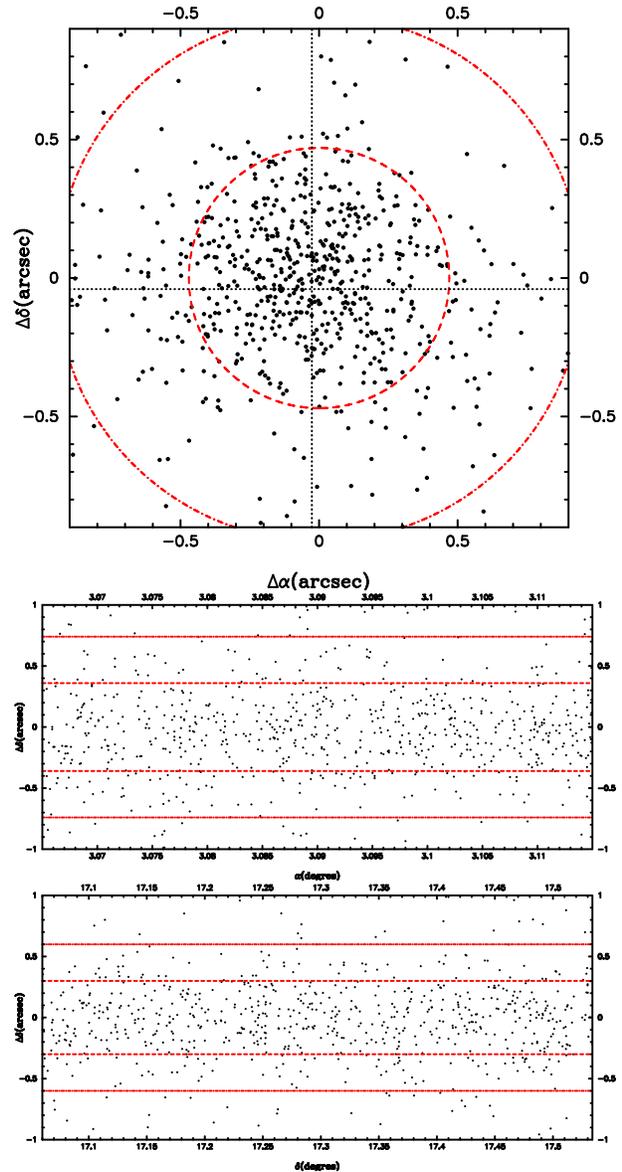

  \includegraphics[width=8cm]{graph_deltaRA_deltaDEC.PGPLOT.LW5.ps}
  \includegraphics[width=8cm]{graph_RA_deltaRA.PGPLOT.LW5.ps}
  \includegraphics[width=8cm]{graph_DEC_deltaDEC.PGPLOT.LW5.ps}
  \caption{Astrometric calibration of the MS0302+17 field. The top
    panel shows the residual $\Delta \alpha$ and $\Delta \delta$ between
    the USNO reference star position and the astrometric solution. The two bottom
    panels show the $RA$ and $DEC$ residual star position difference as a function
    of the $RA$ and $DEC$ in the MS0302+17 supercluster field. No trend is
    visible, showing that astrometry is stable across the whole
    field of view.}
  \label{fig:astrometricplot}
\end{figure}

\subsection{Astrometric calibration}\label{subsec:astrometry}
The astrometric calibration was done using the {\tt Astrometrix} package
developed jointly by the TERAPIX center and Osservatorio di Capodimonte
in Naples for wide field images\footnote{\url{http://www.na.astro.it/~radovich/}}.
The algorithms are extensively described in McC03 and \citet{Radovich04},
so we refer to these two papers for further details.

For the MS0302+17 data, the calibration was done using the USNO-A2.0
reference star catalogue \citep{Monet98}. The astrometric center and
tangent point of the CFH12K is $\alpha_{\rm 2000}=\coordra{03}{05}{25.8}$ and
$\delta_{\rm 2000}=\coorddec{+17}{17}{54}$. Since the $V$ image has much fewer
saturated stars than the $R$ image, we preferred to use it as the reference
astrometric data, although the $R$ is deeper and has a better seeing.
We then cross-correlated the
$V$ input catalogue generated by {\tt SExtractor} to the USNO-A2.0. After rejection
of saturated objects, we found 731 stars common to both catalogues.
The star sample is homogeneously spread over the CCDs, so that we used
50 to 70 stars per CCD. \reffig{fig:astrometricplot} shows the residuals
between the reference USNO-A2.0 star positions and positions derived
from the astrometric solution.
The $rms$ coordinate error is $ 0\farcs45$ (68\%) and is
similar in both $RA$ and $DEC$ directions ($\Delta \alpha = 0\farcs36$,
$\Delta \delta =0\farcs30$, respectively). No systematic shift or position
gradient is visible. Each $V$ frame was resampled according to the astrometric
calibration and then stacked to produce the final $V$ image.

The $B$ and $R$ images are calibrated with respect to the $V$ image.
The matching uses detection catalogues generated by {\tt SExtractor}
so the cross-correlation can be done using several thousands
of stars and galaxies. Since the $B$, $V$ and $R$ data were obtained during
the same night, systematic offsets only correspond to small shifts imposed
by the observer, and rotation between each image is negligible. 
 Across such a small field, the atmospheric differential refraction
for an airmass ranging between 1.0 to 1.5 produces shifts smaller
than $0\farcs4$ between the $B$, $V$ and $R$ image and a chromatic
residual that is less than $0\farcs04$ between the $B$ and $R$ image.
The $rms$ coordinate errors between the $B$ and $R$
catalogue and the $V$ reference are $ 0\farcs05$ (68\% CL).
\begin{figure*}
  \centering
  \includegraphics[width=18cm]{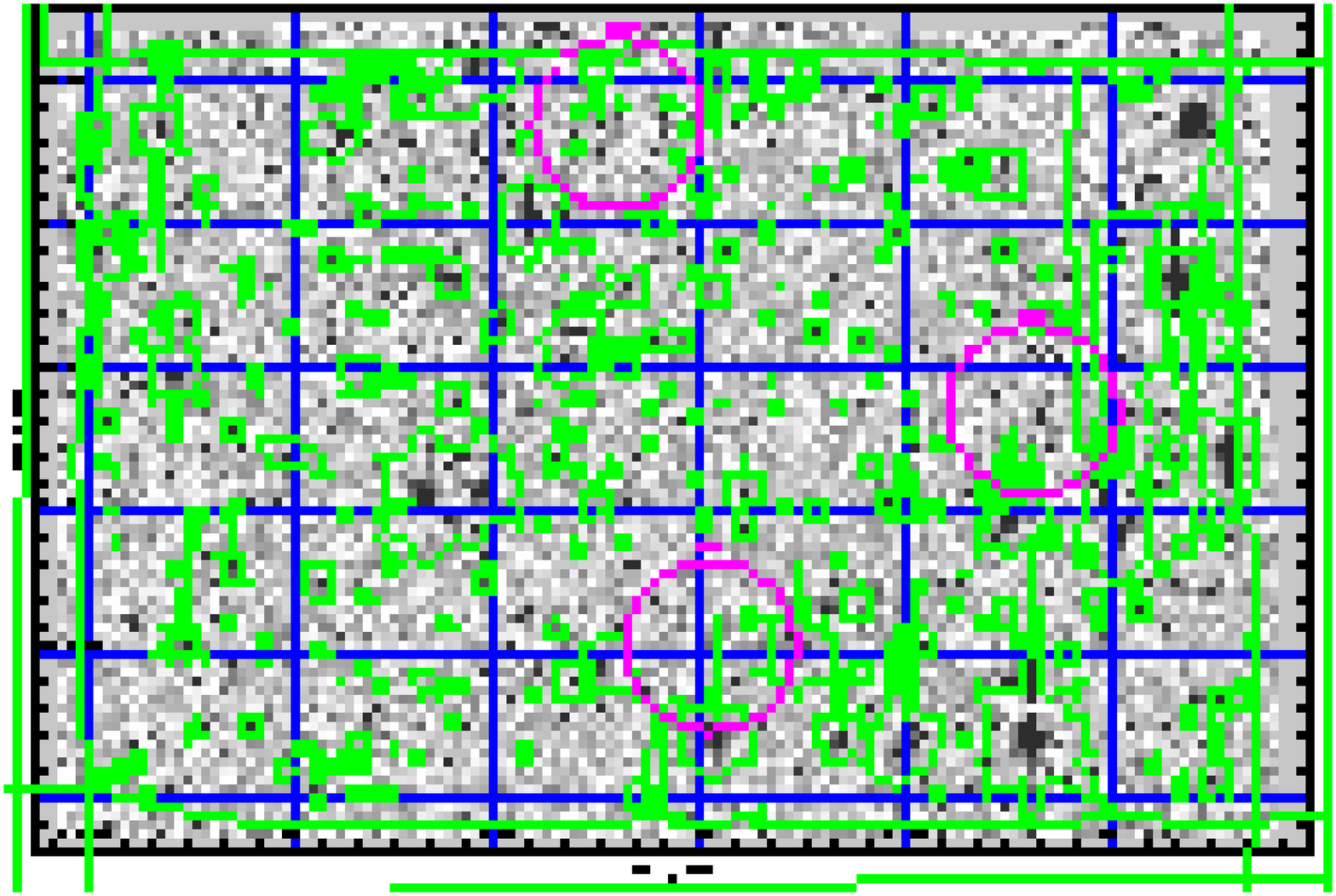}
  \caption{$R$ band image of the MS0302+17 supercluster field used
    for the light and mass analysis. Coordinates are in J2000.
    The center point is $RA=\coordra{03}{05}{25.79}$ and
    $DEC=\coorddec{+17}{17}{54.02}$.
    The field size is $42.9\arcmin\times29.0\arcmin$
    ({\sl i.e.} $14.2 \hmMpc \times 9.6 \hmMpc$ @ $z=0.42$).
    The 3 arcmin radius circles show the cluster
    positions of the X-ray emission peaks:
    ClN, North $\coordra{03}{05}{17.81},\,\coorddec{+17}{28}{37.6},\, z=0.425$,
    ClE, East $\coordra{03}{06}{18.89},\,\coorddec{+17}{18}{33.9},\, z=0.418$ and
    ClS, South $\coordra{03}{05}{31.49},\,\coorddec{+17}{10}{16.3},\, z=0.426$.
    Green polygons define the masked areas that have been removed from the 
    image analysis.}
  \label{fig:ms0302BVRfield}
\end{figure*}

\subsection{Final catalogues}\label{subsec:catal-summary}
Using the calibrated $B$, $V$, and $R$ images, we then produced
the final $BVR$ objects catalogue. It contains 28600 galaxies and
1100 stars, as defined according to the {\tt SExtractor} stellarity index.
The common area is composed of 12500 $\times$ 8500
pixels and is 0.343 deg$^2$ wide. It corresponds to an angular
size of $42\farcm7 \times 28\farcm9$
({\sl i.e.} $14.2 \hmMpc \times 9.6 \hmMpc$ at $z=$0.42)\ .

For weak lensing analysis we produced another catalogue that only uses
the  $R$ band image without regard of the $B$ and $V$ data. This image 
is the deepest one and has the best seeing. The weak lensing catalogue
contains more objects than the joined $B$, $V$ and $R$ one.
Its properties are detailed in \refsec{sec:weak-lensing}.

CCD defects, gaps or overlap areas between CCDs, bright stellar halos,
saturated stars and asteroid track residuals generate spurious features.
They are removed from the catalogue using the manual masking process described
in \citet{Waerbeke00, Waerbeke01} as shown in \reffig{fig:ms0302BVRfield}.
Field boundaries are also masked, and we finally end up with an effective
area of 0.228 deg$^2$. The catalogues discussed in the following will only
concern this common unmasked part of the field.

\section{Supercluster galaxies and light distribution}\label{sec:cluster-photom}
As shown above, the color information is stable across the field
within 0.05 mag accuracy which is sufficient to make a reliable selection
of cluster and non-cluster galaxies using colors.
The supercluster member selection and the redshift distribution of foreground
and background lensed sources were done using color-color diagrams
together with the measurement of photometric redshifts.

\subsection{Photometric Redshifts \vs Color-Color relation}\label{subsec:photoz-colmag}
We first attempted to use $B$, $V$ and $R$ photometry to derive photometric 
redshifts using the
{\tt hyperz}\footnote{\url{http://webast.ast.obs-mip.fr/hyperz/}}
package \citep{Bolzonella00}. Details of the method applied to clusters or
deep multi-color wide field surveys
can be found in \citet{Athreya02,Waerbeke02,gavazzi03}.
However, compared to these previous analyses, we only have three bands,
which severely reduces the reliability of photometric redshift information.
When compared to an ``empty'' region located westward on the field,
the photometric redshifts show an excess of galaxies in the redshift range
$\left[0.25-0.65\right]$. The photometric redshift uncertainty ($\sim\pm0.2$)
hampers any detailed redshift investigation of the supercluster galaxies.
This noisy redshift information can partially be used for the distinction
between foreground and supercluster objects and background lensed galaxies
(See \refsec{sec:weak-lensing}).

\begin{figure}[tbh]
  \centering
  \includegraphics[width=8.5cm]{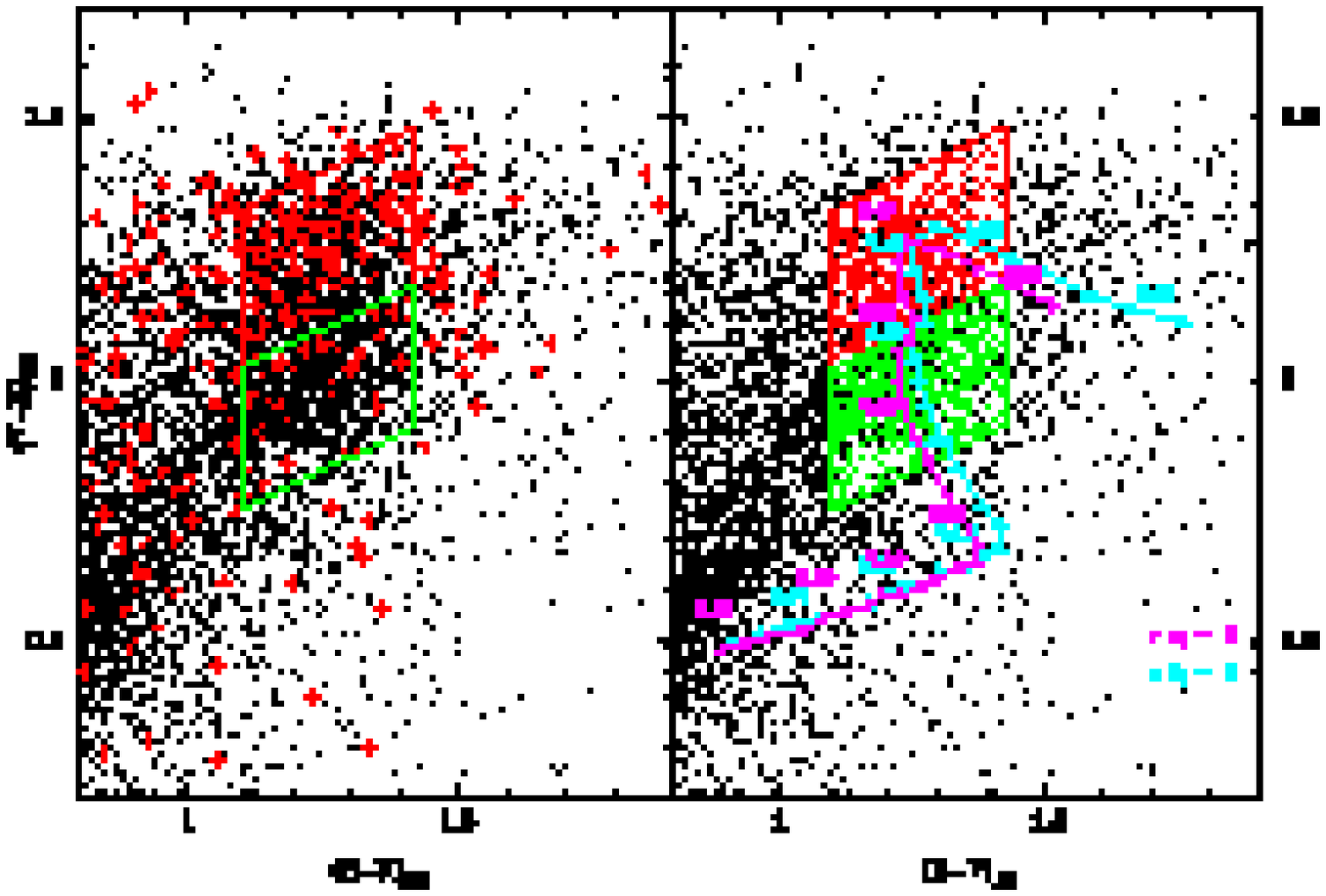}
  \includegraphics[width=8.5cm]{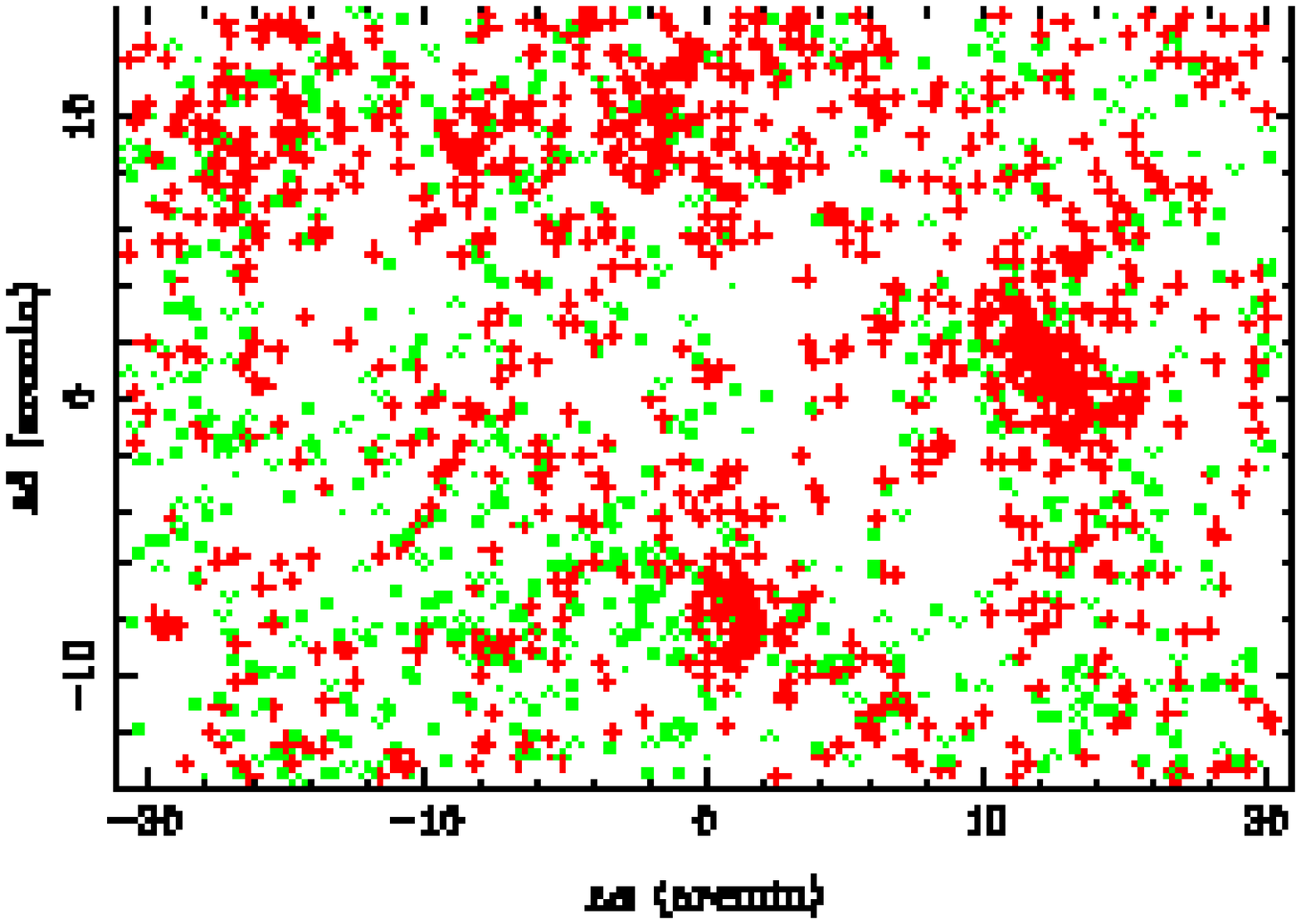}
  \caption{Upper left panel: Color-color diagram. The red + signs code
    for the supercluster objects that lie within 1.5 arcmin of individual
    clusters. The early-types are concentrated around $(B-V)_{AB} \approx 1.2$.
    The additional knowledge of $(V-R)$ color provides
    an efficient selection of supercluster objects at $z\sim0.4$.
    \ \ \ 
    Upper right panel: Color-color diagram that explicitly makes the distinction
    between supercluster and foreground ellipticals along the $(V-R)$ axis.
    The variation of $(V-R)$ with redshift along the $(B-V)_{AB}\approx1.2$ line
    is well reproduced by predicted passive evolution tracks for ellipticals
    whatever their formation redshift. We define a foreground subsample
    centered around $z \approx 0.3$ {\sl i.e.} inside the lower green lozenge
    whereas the upper red one encompasses the supercluster ellipticals.
    \ \ \
    The lower panel shows the corresponding spatial distribution. Supercluster
    members (red + symbols) are clearly clustered around the three known clusters.
    Foreground objects (green $\times$ symbols) are clearly less clustered
    although some clumps are also visible.}
  \label{fig:colmagdiag}
\end{figure}

Fortunately, at redshift $z \approx 0.4$ the typical 4000\AA~ break spectral
feature lies between the $B$ and $R$ filters and can easily reveal early type 
galaxies.
The cluster selection was therefore primarily focused on the red cluster
sequence in the color-magnitude diagrams. By using a $B$, $V$ and $R$
color-color diagram of well-defined magnitude limited sample 
of galaxies for the selection the supercluster members, 
one can easily isolate co-eval early-type cluster galaxies. We first
flagged objects within a 1.5 arcmin radius from the center of each cluster.
\reffig{fig:colmagdiag} shows the $(B-V)$ versus $(V-R)$ color-color diagram.
Early-type galaxies concentrate around $1.1\le(B-V)\le1.42$, the supercluster
elliptical galaxies at $z=0.42$ having also $0.47\le(V-R)-0.5(B-V)\le0.77$.
Objects with a $(V-R)$ color bluer than that band are elliptical galaxies
with a lower redshift.  Most are in the redshift range 
$0.1 < z < 0.4$. Passive evolution tracks for elliptical galaxies kindly
provided by D. Leborgne \citep[see][]{fioc97} are in excellent agreement with
our observations.

The following set of equations summarizes our selection criteria for
early-type galaxies. Equations \eqref{eq:colmagrela}-\eqref{eq:colmagrelc}
are criteria for supercluster members, whereas    
\eqref{eq:colmagrela}-\eqref{eq:colmagrelb} and \eqref{eq:colmagreld}
stand for the selection of foreground early-type galaxies.
\begin{subequations}\label{eq:colmagrel}
  \begin{gather}
    19\,\le\,R\,\le\,23,\label{eq:colmagrela}\\
    1.1\,\le\,(B-V)\,\le\,1.42\label{eq:colmagrelb}\\
    0.47\,\le\,(V-R)-0.5 (B-V)\,\le\,0.77\label{eq:colmagrelc}\\
    0.2\,\le\,(V-R)-0.5 (B-V)\,\le\,0.47\label{eq:colmagreld}
  \end{gather}
\end{subequations}

The foreground sample has a mean redshift $z \simeq 0.3$ and
contains 770 galaxies, and the supercluster sample contains 750 galaxies.
Their luminosity distribution is well centered around $L_*$.
We did not select ellipticals with $z<0.2$ since they are very few and
provide a negligible lensing signal. Elliptical galaxies at $z\ge0.6$ also
have a poor lensing efficiency compared to those at the supercluster 
redshift.

Our color-color selection method fails to localize bluer late-type galaxies.
Thus, we have to keep in mind that the light due to cluster spirals is not taken
into account. K98 and \citet{fabricant} argued that $\approx 30\%$ of the 
total $B$ band rest-frame luminosity is due to late-types, so  
their contribution, though sizeable, is not expected to be dominant.

We estimated the contamination by field galaxies inside the
color-color region of supercluster members. We selected similar
galaxies satisfying \eqref{eq:colmagrela}-\eqref{eq:colmagrelc}
far from the regions around the three clusters and
plotted them in the color-color diagram. The fraction of field galaxies
inside the cluster color-color region turns out to be negligible.
This efficient selection process expresses the fact that
for this redshift $(z \approx 0.4$), the $(B-V)$ and $(V-R)$ colors
are reliable filters. In contrast, the lower $(V-R)$ limit used for the
foreground subsample selection is more questionable, so we may miss some
of the nearest ellipticals. However, since their lensing contribution
is small, it has no impact on the interpretation of the weak lensing signal.

\subsection{Spatial distribution of light \& associated convergence}\label{subsec:light-distrib}
In this section, we investigate the spatial distribution of supercluster
galaxies. We also estimate the foreground contribution to light.
The apparent $B_{AB}$ magnitudes are converted into rest-frame luminosities
using K-corrections also provided by D. Leborgne. For $z=0.42$
supercluster members, we used $K_B = 1.88$, and for $z\simeq0.3$
foreground early-types $K_B = 1.14$. As detailed in \refsec{subsec:sigma-crit},
the redshift distribution of lensed sources implies
the following critical surface densities:
$\scrit(z_d=0.42) = 2.72\times10^{14}\hmMsun/{\rm arcmin}^2$ and
$\scrit(z_d=0.3)  = 1.95\times10^{14}\hmMsun/{\rm arcmin}^2$.

Instead of computing a filtered luminosity map with Gaussian smoothing, we
follow the method of K98, G02 and \citet{wilson01}. It consists of weighting
the luminosity map by its lensing efficiency (different for foreground and
supercluster components). Assuming a constant mass-to-light ratio for
supercluster galaxies, the luminosity field at a given position
$\vec{\theta}$ is converted into a mass density field,
$\Sigma_{\mathcal L}(\vec{\theta})$. Then it is translated into a convergence
$\kappa_{\mathcal L}(\vec{\theta})=\Sigma_{\mathcal L}(\vec{\theta})/\scrit$
field. We finally convert it into a shear field $\gamma(\vec{\theta})$
and derive a supercluster shear pattern that samples the field according to the
source catalogue positions (see \refsec{subsec:sigma-crit}).
From this shear field a new convergence field can be drawn. It has the
same field size and shape, the same masking and the same sampling properties
as the $\kappa$-map we construct in \refsec{subsec:massmap}.

The inferred shear field reads :
\begin{equation}\label{eq:lum2shear}
  \gamma(\vec{\theta}) = \left[\sum_{i=1}^{N_{\rm lens}} \gamma_0(\vec{\theta}-\vec{\theta_i})\right] * W(\vec{\theta}),
\end{equation}
where $W$ is a 40 arcsec Gaussian smoothing filter, and $\gamma_0$ is the shear
profile of an individual galaxy. Since we are primarily interested in
the collective behavior of galaxies and we are looking at scales $\ge 1$ arcmin,
we do not make a further hypothesis about the radial density profiles of
galaxy halos. This means that galaxies are equivalent to point masses, so
$\gamma_0$ simply reads:
\begin{equation}\label{eq:shear-ptmass}
  \gamma_0(\theta) = \frac{M}{L} \times \frac{L}{\pi \scrit \theta^2}
\end{equation}
where $M/L$ is the mass-to-light ratio which is assumed to be the same for all
galaxies. The $\gamma \rightarrow \kappa$ inversion is detailed in 
\refsec{subsec:massmap}. It turns out that luminosity-weighted $\kappa_{\mathcal L}$
and number-density-weighted\footnote{in this case, $\gamma_0(\theta) = 
  \frac{M_0}{\pi \scrit \theta^2}$ with $M_0$ the mass of a galaxy halo, which
  is assumed to be constant from one galaxy to another.} 
$\kappa_N$ convergence maps are almost proportional. A small
deviation from equality appears for the highest contrast values. In this case,
$\Delta \kappa_N/\kappa_N \lesssim \Delta \mathcal{\kappa_L}/\mathcal{\kappa_L}$
as we expect if the brightest galaxies lie in the densest regions.
\begin{figure}[tbh]
  \centering
  \includegraphics[width=9cm]{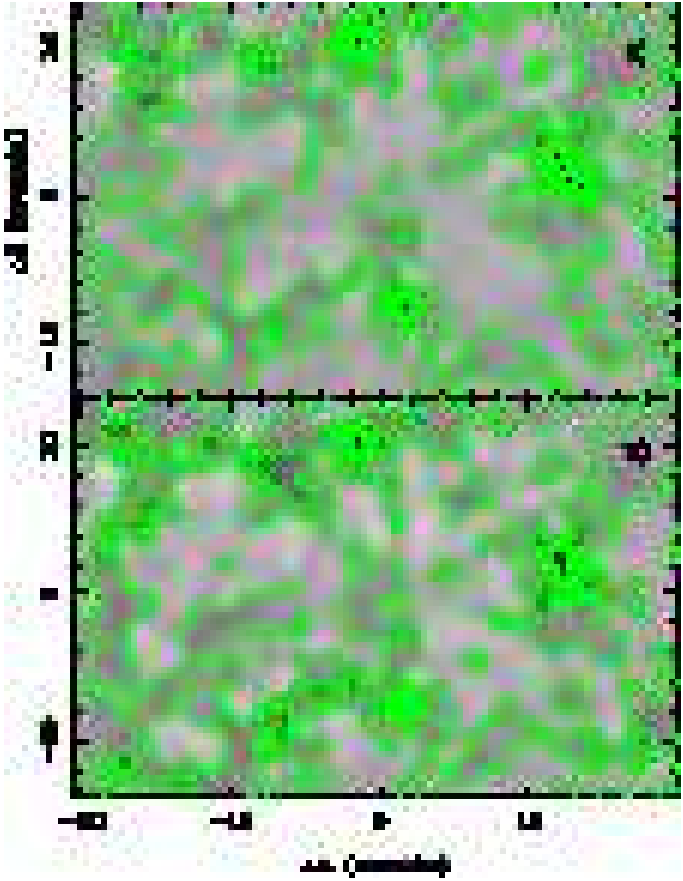}
  \caption{Effective convergence maps derived from the luminosities of
    early type galaxies. The convergence is expressed as
    $\kappa(\vec{\theta})=\Sigma(\vec{\theta})/\scrit$ but the 
    supercluster contribution may be contaminated by foreground galaxies.
    Panel a) (top) shows the luminosity weighted convergence map of 
    supercluster objects, as inferred from the positions of galaxies in 
    the color-color diagram. Panel b) (bottom) shows 
    the same map with the contribution of the foreground galaxies.
    The origin coordinate is
    $RA=\coordra{03}{05}{25.79}$ and $DEC=\coorddec{+17}{17}{54.02}$.
    We applied a $40\arcsec$ wide Gaussian smoothing scale.
    The two maps are almost similar, confirming 
    that foreground structures along the line of sight do not 
    dominate uncertainties in the error budget.
    Green levels start at $\kappa=0.0$ and increase linearly with step $0.02$.
    We assumed a fiducial mass-to-light ratio $M/L=300\mslsun$.}
  \label{fig:light-distrib}
\end{figure}
The resulting $\kappa$ from light maps are shown in \reffig{fig:light-distrib}.
The upper panel shows the $\kappa_\mathcal{L}$ convergence map
for the supercluster objects only, whereas the lower panel shows the modifications
produced by addition of foreground structures. The three known clusters
are clearly detected and seem to encircle a large underdense region. A diffuse
extension, less dense than the clusters, appears westward from the northern cluster
(ClN). This extension encompasses two clumps at $\Delta\alpha \sim -8 \arcmin$
and $\Delta\alpha \sim -15\arcmin$. Another extension toward the North-West
of ClS is partly due to foreground structures. Using the spectroscopic redshift
of two member galaxies, K98 argued that this clump probably lies at $z=0.3$\ .

\section{Weakly Lensed objects sample - Shear analysis}\label{sec:weak-lensing}
The coherent stretching produced by the weak lensing effect due to the MS0203+17
supercluster is measured using the deep catalogue extracted from the $R$ band 
image. Its depth and high image quality allow us to lower the detection
threshold and to increase the galaxy number density ($\approx$ 25 arcmin$^{-2}$), 
compared to the $B$ and $V$ images.
This reduces the Poisson noise of the weak lensing statistics and
increases the spatial sampling of the supercluster mass reconstructions.
Close galaxy pairs with angular separations
less than 5 arcsec are discarded to avoid blended systems that
bias ellipticity measurements. The reliability of shape measurements is 
expected to be as good as the current cosmic shear survey data 
\citep{Waerbeke01}.

\subsection{PSF correction}\label{subsec:psf-cor}
Blurring and distortion of stars and galaxies produced by
instrument defects, optical aberrations, telescope guiding,
atmospheric seeing and differential refraction are corrected
using the PSF of stars over the whole field. Several correction techniques
and control of systematic errors have been  proposed over the past
10 years \citep[see \eg][]{Mellier99,BartShneid01,WaerbMell03,Refregier03}.
In the following we use the most popular KSB95 method
initially proposed by \citet{KSB95}.
Several teams have already demonstrated that the
KSB95 method can correct systematics residuals down to the 
lower limit  shear amplitude expected on supercluster scales
\citep{WaerbMell03,Refregier03}.

\begin{figure}[tbh]
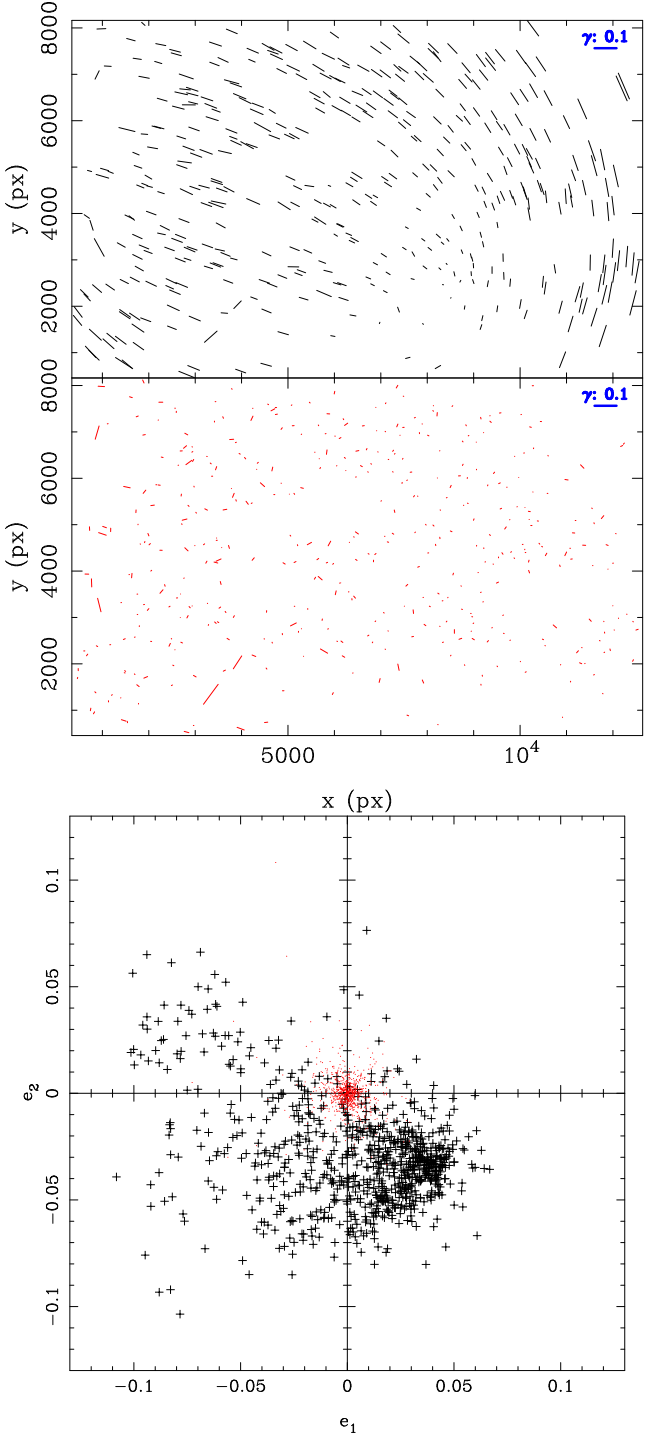

  \centering
  \includegraphics[width=8.5cm]{psf_anis1.eps}
  \includegraphics[width=8cm]{psf_anis2.eps}
  \caption{Upper panel :  Spatial distribution of stellar ellipticities
    before (black) and after (red) PSF correction.
   \ \ \  Lower panel :  The same stellar ellipticities in the $(e_1,e2)$ plane, 
    uncorrected (black crosses) and corrected (red dots)
    from PSF anisotropy. The residual $rms$ dispersion around the center
    is less than one percent.}
  \label{fig:psf-anis}
\end{figure}

Following KSB95 method, the observed ellipticity components
$ e^{\rm obs}_{\alpha=1,2}$ are composed of its intrinsic ellipticity components
$e^{\rm src}_\alpha$, and linear distortion terms that express the instrument
and atmospheric contaminations and the contribution of
gravitational shear to the galaxy ellipticity. Each ellipticity component
is transformed as:
\begin{subequations}\label{eq:KSB}
  \begin{gather}
    e^{\rm obs}_\alpha = e^{\rm src}_\alpha + P^{\rm g}_{\alpha\beta} g_\beta -
    P^{\rm sm}_{\alpha\beta}q^*_\beta,\label{eq:KSBa}\\
    {\rm with\ \ } P^{\rm g}_{\alpha\beta} = P^{\rm sh}_{\alpha\beta}
    - P^{\rm sm}_{\alpha\gamma} \left(\frac{P^{\rm sh}}{P^{\rm sm}}\right)_
    {\gamma\beta}^*,\label{eq:KSBb}
  \end{gather}
\end{subequations}
where $g$ is the reduced gravitational shear, $P^{\rm sm}$ is the
{\sl smear polarizability}, $P^{\rm sh}$ the {\sl shear polarizability}
and $P^{\rm g}$ the isotropic circularization contribution to
the final smearing. In the following, all these tensors are simplified
to half their trace and have been calculated with
{\tt Imcat}\footnote{\url{http://www.ifa.hawaii.edu/~kaiser/imcat/}} tools.
$\left(\frac{P^{\rm sh}}{P^{\rm sm}}\right)^*$ and $q^*$ are quantities that
are measured from field stars. Their shape is fitted by a second order polynomial,
applied individually to each CCD of the CFH12K camera. Stars are selected in the
magnitude-$r_h$ plane, as usual. $q^*$ is the anisotropic part of the PSF, which
is subtracted from galaxy ellipticities.
The residual is shown in \reffig{fig:psf-anis}\ . It does not show any 
peculiar spatial pattern and is consistent with a one percent $rms$ noise.

The smearing part of the PSF contained in the $P^{\rm g}$ term depends
on the magnitude of the object and on its size as compared to the seeing disk.
To optimally extract $P^{\rm g}$, we derived it from an averaged value over its
70 nearest neighbors in the magnitude$-r_h$ plane.
Its variance is then used as a weighting
scheme for the shear analysis. The weight assigned to each galaxy is
finally the inverse variance $\sigma_{e_i}^2$ of the observed ellipticities :
\begin{equation}\label{eq:weight_scheme}
  w_i = \frac{1}{\sigma_{e_i}^2} = \frac{P^{g2}}
  {P^{g2}\sigma_0^2 + \sigma_i^2},
\end{equation}
where $\sigma_0=0.3$ is the intrinsic dispersion in galaxy
ellipticities and $\sigma_i\approx 0.13$ is the observed dispersion of
ellipticities over the 70 closest neighbors.
A weighted magnitude histogram of the source sample  shows
that we can select galaxies down to the limiting magnitude $R_{AB}=25.6$.
In this subsample, galaxies with a $r_h$ smaller than that of stars are discarded.

\subsection{Redshift distribution \& shear calibration}\label{subsec:sigma-crit}
Because the source sample is deeper than the photometric catalogue,
the redshift distribution of background objects cannot be derived
from the photometric redshifts calculated with the $B$, $V$ and $R$ data.
Nevertheless, for the distinction between foreground and background-lensed
objects, we can use a rough limiting redshift estimate. 

We chose to reject objects
with a $z_{\rm phot} < 0.5$. We also discarded the early-type galaxies
selected in the color-color diagram of \refsec{sec:cluster-photom}.
Finally, we selected background galaxies within the magnitude range :
$22.4<R_{AB}<25.6$. The source catalogue contains 22125 galaxies. This 
corresponds to a number density $n_{\rm bg} \approx 27\,{\rm arcmin}^{-2}$.

The properties of the resulting sample are roughly comparable with those of
the sample of \citet{Waerbeke01,Waerbeke02}, though their magnitude
cut $I_{AB}< 24.5$ instead of $R_{AB} \lesssim 25.6$ as in this work.
They inferred the redshift distribution:
\begin{equation}\label{eq:dist-z}
  n(z)=\frac{1}{z_s \Gamma(a)}
  \left(\frac{z}{z_s}\right)^{a-1} \eexp{-z/z_s},
\end{equation}
where $a=5/2$, $z_s=0.44$ leading to $\bar{z}=a z_s=1.1$ and
$\sigma_z = \sqrt{a}z_s = 0.7$. The median redshift is well 
approximated by ${\rm Med}(z)\simeq(a-0.33)z_s\simeq0.95$.
At the same time, G02 proposed a median redshift $z=1$
for their magnitude cut $R < 26$. Using the same analytic form
as \eqref{eq:dist-z}, we found that $a\simeq 1.9$ and $z_s\simeq 0.55$ provide
a good description of our redshift distribution implying a median redshift
${\rm Med}(z)=1$ and a broader distribution $\sigma_z \simeq 0.8$\ .\\
For the supercluster redshift $z_d=0.42$ we calculated the mean of the ratio
$\beta=\left<D_{ds}/D_s\right>$ and the corresponding critical surface density
$\scrit = \frac{c^2}{4\pi G} \frac{\beta^{-1}}{D_d}$, where
$D_d$, $D_s$ and  $D_{ds}$ are angular distances between the observer and deflector,
observer and sources and deflector and sources, respectively. We found :
\begin{equation}\label{eq:sigma-crit}
  \begin{split}
    \beta & = 0.49,\\
    \scrit &= 2.72 \times 10^{15}\,h_{70}\,
    {\rm M}_\odot\,{\rm  Mpc}^{-2}\\
    &=3.02 \times 10^{14}\,\hm\,{\rm M}_\odot\,{\rm  arcmin}^{-2}.
  \end{split}
\end{equation}
For foreground galaxies at $z \sim 0.3$, we found $\beta=0.63$,
$\scrit = 1.95 \times 10^{14}\,\hm\,{\rm M}_\odot\,{\rm  arcmin}^{-2}$.
The redshift distribution of sources is indeed equivalent to a single
source plane configuration with redshift $z_{\rm sheet} \approx 0.95$.
The depth and the source plane redshift we use are in good
agreement with previous ground based analyses like those of
\citet{clowe01,clowe02}. The uncertainty in the gravitational
convergence produced by the redshift distribution of the sources is 
about 5\%, which is much smaller than the error bars we expect
from statistical noise due to intrinsic galaxy ellipticities. 

\subsection{Mass Map}\label{subsec:massmap}
\begin{figure}[tbh]
  \centering
  \includegraphics[width=8.cm]{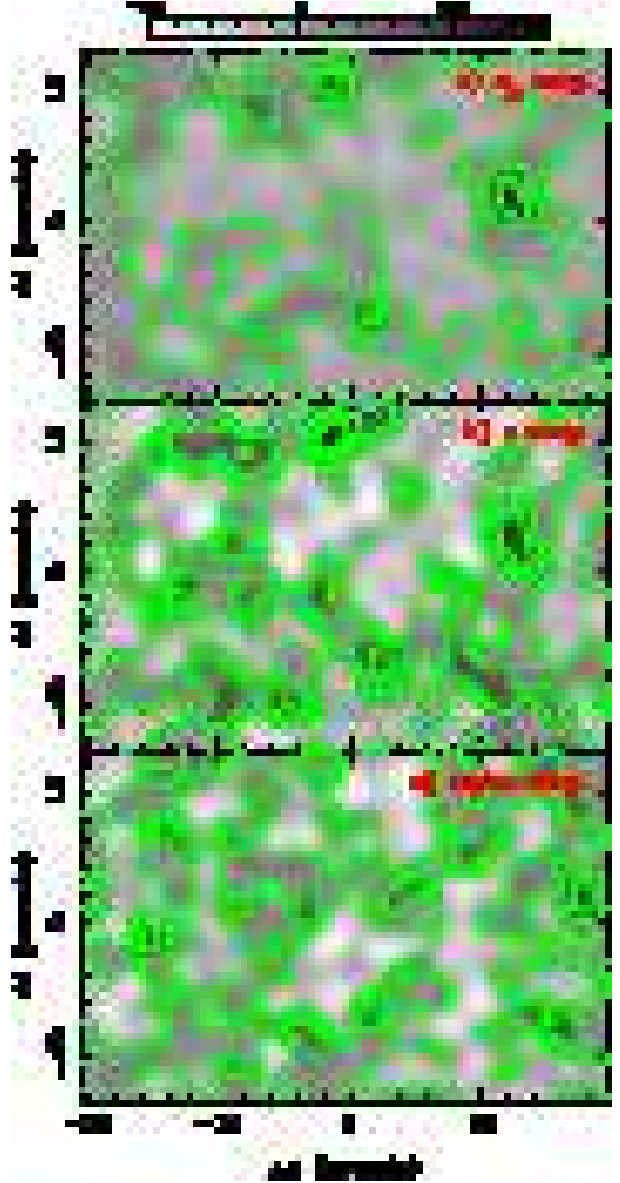}
  \caption{Upper panel: Mass reconstruction derived from the 
    light distribution emitted by clusters+foreground elliptical galaxies
    $\kappa_L$. This map is almost a reproduction of the b) panel
    of \reffig{fig:light-distrib} with one arcminute filtering scale.
    Middle panel: Reconstructed dark matter surface density
    $\kappa(\vec{\theta})$ using the KS93 inversion technique.
    Lower panel: Same reconstruction after $45\degr$ rotation of source galaxies.
    No noticeable patterns due to  systematics are visible. 
    In all maps, the shear has been sampled at the observed
    position of background source galaxies, and the maps suffer the same
    (edge+mask)-effects. The Gaussian smoothing scale is $1\arcmin$.
    Levels are the same as in \reffig{fig:light-distrib}.
    The overall agreement between panels a) and b) is good.}
  \label{fig:kappamap}
\end{figure}

Our mass reconstruction is based on the \citet{KS93} (KS93) algorithm.
The convergence $\kappa(\vec{\theta})=\Sigma(\vec{\theta})/\scrit$
is related to the observed shear field $\gamma(\vec{\theta})$ through:
\begin{equation}\label{eq:kappa2gamma}
  \kappa(\vec{\theta})=\int_{\mathbb{R}^2} D(\vec{\theta}-\vec{\vartheta})^* \gamma(\vec{\vartheta}) \der^2 \vec{\vartheta},
\end{equation}
where $\gamma$ and $D(\vec{\theta})=\frac{1}{\pi}\frac{-1}{(\theta_1-i \theta_2)^2}$
are complex quantities. On the physical scales we are exploring the lensing
signal is weak enough so that 
$\langle e \rangle = \frac{\gamma}{1-\kappa} \simeq \gamma$. The ellipticity
catalogue is smoothed with a $\theta_s=40 \arcsec$ Gaussian filter :
\begin{equation}\label{eq:gamma-estim}
  \hat{\gamma}(\vec{\theta}) = \frac{1}{N} \sum_i w_i\,e_i\:
  \exp\left(-\frac{\left(\vec{\theta}-\vec{\theta}_i\right)^2}{2 \theta_s^2}\right),
\end{equation}
where $w_i$ are the weights defined in Eq. \eqref{eq:weight_scheme} and   
$N\simeq 2\pi n_{\rm bg} \theta_s^2 \approx 170$ can be viewed as the mean number
of sources inside the filter. The resulting convergence map presents correlated
noise properties :
\begin{equation}\label{eq:map-noise-ppties}
  \langle \kappa_n(\vec{\vartheta}) \kappa_n(\vec{\vartheta}+\vec{\theta})
  \rangle = \frac{\sigma_i^2}{8 \pi n_{\rm bg} \theta_s^2}
  \exp\left(-\frac{ (\vec{\theta}-\vec{\vartheta})^2}{4\theta_s^2}\right).
\end{equation}
$\sigma_i\approx0.42$ is the dispersion in ellipticities of our galaxy sample.
$\frac{\sigma_i}{\sqrt{8\pi n_{\rm bg} \theta_s^2}} \simeq 0.016$
characterizes the noise level.
The $\kappa$-map reconstruction result is shown in the middle panel of 
\reffig{fig:kappamap}. The bottom panel shows the reconstruction applied
to the same galaxy sample, but with the orientation rotated by $45\degr$.
It represents the imaginary part of Eq. \eqref{eq:kappa2gamma},
which should be a pure noise realization if the coherent distortion field
is only produced by gravitational lensing.

The three main clusters ClN, ClS \& ClE are detected with a 
high significance. A few substructures with a lower detection 
significance are also visible.
We detail in \refsec{subsec:clust_proper} the various quantities
measured for the clusters. A comparison with the X-ray emissivity map displayed
on Fig. 1 of K98 shows an excellent agreement. See also Fig. 1 of
\citet{fabricant}. The clusters are also detected in the
$\kappa$-from-light map shown in the top panel of \reffig{fig:kappamap}.
The clumpy extension detected westward from ClN can be seen in the light map.
Another clump located at $\coordra{03}{04}{30}$, $\coorddec{+17}{15}{00}$
may be either a foreground structure or an extension toward the west from ClS. 
This luminous component is visible in the lower panel of
\reffig{fig:light-distrib} but not in the $\kappa$-map of the top panel.
From spectroscopic data, K98 argued that this structure may lie at
$z\approx0.3$. A large void region between the clusters is also apparent
in the mass map. When decreasing the smoothing scale, the core of ClE 
splits into two maxima that are also visible in the higher resolution
light map of \reffig{fig:light-distrib}.

\subsection{Properties of clusters}\label{subsec:clust_proper}
The global properties of the three clusters are explored using integrated
physical quantities enclosed within the radius $r_0 = 1 \hmMpc (= 3\arcmin$). 
For each cluster the center is set to the X-ray emissivity center.
Table \ref{tab:summary} summarizes the main quantities : the total mass from
weak lensing estimates (row 3), the total rest-frame B luminosity emitted by
the supercluster early-type galaxy sample (row 4), the inferred mass-to-light
ratio (row 5), the ``spectroscopic'' velocity dispersion compiled from
\citet{Dressler92,fabricant,carlberg} (row 6),
and the velocity dispersion derived from a
fit of the weak lensing data to a Singular Isothermal Sphere (SIS) model 
(row 7): $\kappa(\theta)=\theta_E/2\theta$ , where  
$\theta_E = 4 \pi \left(\frac{\sigma}{c}\right)^2 \beta$ is  the
Einstein radius. Rows (8) and (9) are X-ray  ROSAT HRI/IPC
and ASCA data \citep{gioia94,fabricant,Henry00}\footnote{Possible corrections
  to these values and larger error bars may be found in \citet{ellis02,Yee03}.
  Since the following analysis does not deal with these data, we refer to these
  papers for further information concerning the supercluster's X-ray properties.}.
Row (3) is computed using the densitometric $\zeta-$statistic :
\begin{equation}\label{eq:zeta-stat}
  \begin{split}
    \zeta(\theta,\theta_0) = &\left\langle \kappa (\theta^\prime
      < \theta)\right\rangle- \left\langle \kappa (\theta < \theta^\prime
      < \theta_0)\right\rangle.\\
    = & \frac{2}{1-\left(\theta_0/\theta\right)^2} \int_\theta^{\theta_0}
    \left\langle\gamma_t(\theta^\prime)\right\rangle \der \ln \theta^\prime.
  \end{split}
\end{equation}
$\tilde{M}(\theta)=\scrit \pi (D_d\theta)^2 \zeta(\theta,\theta_0)$
gives a lower bound on the mass contained in the cylinder of radius $\theta$.
$\left\langle\gamma_t(\theta)\right\rangle$ is the average tangential
shear. $\theta_0$ is set to 7 arcmin. In practice we used the estimator
\begin{subequations}\label{eq:zeta-estim}
  \begin{gather}
    \hat{\zeta}(\theta,\theta_0) = \frac{\sum_{i \in I} w_i\,e_{t,i}\left(\frac{\theta_0}{\theta_i}\right)^2}{\sum_{i \in I} w_i}\label{eq:zeta-estima}\\
    {\rm Var}(\hat{\zeta}) = \frac{\sum_{i \in I} w_i^2\,\sigma_i^2\left(\frac{\theta_0}{\theta_i}\right)^4}{\left(\sum_{i \in I} w_i\right)^2}\label{eq:zeta-estimb} \ , 
  \end{gather}
\end{subequations}
with $I=\left\{i \mid \theta<\theta_i<\theta_0 \right\} $.
The SIS $\theta_E$ value is obtained by  a $\chi^2$  minimisation :
\begin{equation}\label{eq:maxlik}
  \chi^2 \approx \sum_i w_i \left( e_{t,i} - \frac{\theta_E}{2\theta_i} \right)^2,
\end{equation}
where $e_t$ is the tangential ellipticity relative to the cluster center.
A trivial estimator for $\theta_E$ is :
\begin{equation}\label{eq:einstein-estim}
  \hat{\theta}_E = 2\frac{\sum_i w_i e_{t,i}/\theta_i} {\sum_i w_i/\theta_i^2}.
\end{equation}
Note however that this estimator is no longer valid when $\kappa\sim1$.
This means that we have to select galaxies far enough from centers of clusters.
Typically, we set $1\arcmin<\theta<7 \arcmin$.

\begin{table*}
  \centering
  \caption{Summary of cluster properties. $\tilde{M}$ is a lower bound
    on the cluster mass. $L_B$ is the rest-frame blue band luminosity.
    $M/L$ is the mass-to-light ratio. These three quantities are calculated
    inside $r_0=1\hmMpc$. $\sigma_{\rm vel}$ is the kinematic velocity dispersion
    compiled from spectroscopic data \citep{Dressler92,fabricant,carlberg}
    whereas $\sigma_{\rm SIS}$ is the velocity dispersion deduced from
    weak lensing when fitting an isothermal profile for the cluster
    dark matter halo. Note the two distinct values of $\sigma_{\rm vel}$ for ClS.
    The lower value from \citet{carlberg} is based
    on a larger galaxy sample. It is also in better agreement with our estimate.
    $L_{\rm X,bol}$ is the bolometric X rays luminosity
    and $T_{\rm X}$ the gas temperature \citep{gioia94,fabricant,Henry00}.}
  \begin{small}
    \begin{tabular}{rlrrr}\hline\hline
        &                    & ClN & ClS & ClE \\\hline
      (1)  & $\alpha_{\rm 2000}$ & $\coordra{03}{05}{18}$ & $\coordra{03}{05}{31}$ & $\coordra{03}{06}{19}$\\
      (2)  & $\delta_{\rm 2000}$ & $\coorddec{+17}{28}{38}$ & $\coorddec{+17}{10}{16}$ & $\coorddec{+17}{18}{34}$ \\
      (3)  & $\tilde{M}(<r_0)$\ \ \ $[10^{13}\,\hmMsun]$ & $33.1\pm6.7$ & $17.9\pm5.0$ & $15.0\pm5.4$\\
      (4)  & $L_B(<r_0)$\ \ \ $[10^{11}\,h_{70}^{-2}\,{\rm L}_\odot]$ & $8.3\pm1.7$ & $6.7\pm1.5$ & $11.5\pm2.0$\\
      (5)  & $M/L_B$\ \ \ $[\mslsun]$ & $398\mypm{136}{101}$ & $266\mypm{113}{85}$ & $130\mypm{57}{49}$\\
      (6)  & $\sigma_{\rm vel}$\ \ \ [km s$^{-1}$] & $821\mypm{137}{94}$ & $646\pm{93}$ ($921\mypm{192}{123}$) & $912\pm{200}$ \\
      (7)  & $\sigma_{\rm SIS}$\ \ \ [km s$^{-1}$] & $817\mypm{83}{107}$& $635\mypm{109}{131}$ & $595\mypm{110}{133}$\\
      (8)  & $L_{\rm X,bol}$\ \ \ $[10^{44}\,h_{70}^{-2}\,{\rm erg~s}^{-1}]$ & $2.75\pm0.31$ & $3.47\pm0.26$ & $1.84\pm0.36$ \\
      (9) & $T_{\rm X}$\ \ \ [keV] & -- & $4.6\pm0.8$ & --\\\hline\hline
    \end{tabular}
  \end{small}
  \label{tab:summary}
\end{table*}

Finally, the rest-frame B-band luminosity is obtained by adding up the 
luminosities of cluster galaxies with increasing radius. Systematics due to
the selection of supercluster members or to contamination dominate
the error budget but are small (of order 5\% when changing the limits of
Eqs. \eqref{eq:colmagrel} by 10\%). To account for cosmic variance,
we increased the Poisson noise error by a factor of 1.3,
as suggested by \citet{longair79}.

The three clusters differ from one another in terms of mass and luminosity.
ClN is the most massive and has the highest mass-to-light ratio.
ClS shows apparent properties of a well relaxed cluster. It is highly
concentrated with strong lensing features between the two brightest 
cluster galaxies \citep{Mathez92} and a rather high X-ray luminosity.
ClE seems more complex: 
it is the most luminous in the $R$-band although it is the least massive and
the least X-ray luminous. Table \ref{tab:summary} shows that its kinematical
velocity dispersion is much higher than what we infer from weak lensing.
The latter estimate is more typical of a cluster mass than the
value derived from kinematic data. Hence, ClE is likely not relaxed. 
We attempted to describe its
bimodal structure (see \reffig{fig:light-distrib}) by fitting two individual
isothermal spheres at the location of the luminosity peaks ClE1
($\coordra{03}{06}{16.5},\coorddec{+17}{21}{18}$) and ClE2
($\coordra{03}{06}{19.9},\coorddec{+17}{18}{21}$). We found
$\sigma_{\rm ClE1}=312\mypm{100}{215} \,{\rm km s}^{-1}$ and
$\sigma_{\rm ClE2}=473\mypm{84}{100} \,{\rm km s}^{-1}$. The fit quality is
slightly improved, though the quadratic sum of these individual velocity
dispersions is comparable to the single isothermal sphere fit in table
\ref{tab:summary}. Note that ClE is at $z=0.418$ which is a rather high
radial distance to the other clusters. The previous studies of \citet{fabricant}
and K98 demonstrated that ClE might not be gravitationally bound
to the supercluster system.\\
The X-ray luminosity presents a better correlation with mass than with B-band
luminosity. The mass-to-light ratios are rather different but the mean value within
1 megaparsec is $M/L = 249 \mypm{41}{32} \mslsun$. Within $500\hmkpc$ we found
$M/L = 231\mypm{60}{47}\mslsun$ showing that no significant variation with radius
is observed. It is worth noticing that values of $M/L$ for individual clusters
have a larger scatter.\\
We found larger errors than K98 for $\tilde{M}$, but our
estimates are not based on smoothed mass maps from randomly shuffled catalogs.
We directly used galaxy ellipticities in Eq. \eqref{eq:zeta-estim}.
Hence, our error estimates are more conservative and do not suffer
edge + smoothing effects (+ uncontrolled residual correlations).

\section{Correlation Analysis}\label{sec:correlation}
\subsection{Linear biasing hypothesis}\label{subsec:lin-bias-cor}
The high signal to noise ratios and the good resolutions of the light
and mass maps are sufficient to explore how light and mass correlate
and how these quantities evolve as a function of angular scale.
The statistical properties of the relation between dark and luminous matter 
components can then be analyzed from the cross-correlation of 
the $\kappa$ mass map with the $\kappa$-from-light map
shown in panels b) and a) of \reffig{fig:kappamap}.

Let us first assume a simple linear relation between the
luminosity from early-type galaxies (cluster+foreground) and the dark matter
component. The construction of the $\kappa_E$ map for the luminosity
of early-type galaxies is detailed in \refsec{subsec:light-distrib}.
We compute ``light'' maps again
by adopting the same scaling relation as in Eq. \eqref{eq:shear-ptmass}
with a starting mass-to-light ratio $ M/L = 300\mslsun $.
The linear biasing hypothesis between the dark matter convergence fields
$\kappa_M$ and $\kappa_E$ simply reads :
\begin{equation}\label{eq:general-model}
  \kappa_M = \,\lambda\, \kappa_{E}.
\end{equation}
Hence, $300 \lambda$ is the mean mass-to-light ratio. If we assume that it is
constant with scale and redshift, $\lambda$ is easily constrained by
the cross-correlation analysis.

\begin{figure*}
  \includegraphics[width=18cm]{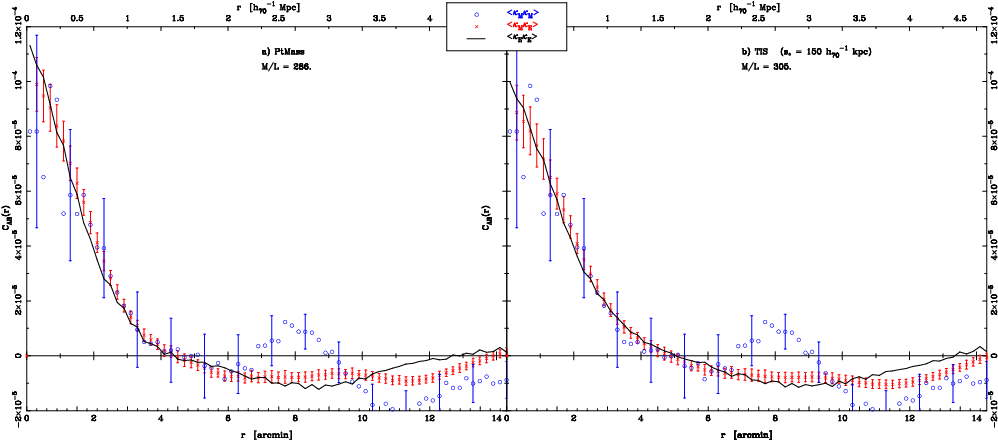}
  \caption{Left panel a) : \ \ $\mycovv{M}$, $\mycov{M}{E}$ and $\mycovv{E}$
    correlation functions for a mass-to-light ratio $ M/L = 286\pm36\mslsun$
    that fits the correlation functions at scales $\theta\lesssim 8 \arcmin$.
    \ \ Right panel b) : \ \ Same plot with a TIS halo model (with truncation radius
    $s_*=150\hmkpc$). With the same assumption $M\propto L$ and a slightly lower
    $\chi^2$, this model also confirms the general conclusion
    {\sl ``light traces mass''}, provided the truncation radius
    $s_* \lesssim 200 \hmkpc$. Note that the bins are correlated.
    For clarity, in both cases one fifth of the error bars is displayed
    for $\mycovv{M}$. This coarser sampling roughly shows
    the required spacing for independent bins.}
  \label{fig:correl}
\end{figure*}

We compute the two-dimensional and azimuthally averaged cross-correlations:
\begin{equation}\label{eq:correl-def}
  C_{AB}(\vec{\theta}) = \langle \kappa_A(\vec{\vartheta}) \kappa_B(\vec{\vartheta}+\vec{\theta}) \rangle \equiv \mycov{A}{B}.
\end{equation}
We have to subtract the noise contributions to the correlation functions.
Since noise properties of mass and light are not correlated,
we only have to calculate the noise autocorrelations :
\begin{equation}\label{eq:noise-subtraction}
  \mycov{A}{A} \longrightarrow \mycov{A}{A} - \mycov{A}{A}_{\rm noise}.
\end{equation}
Noise autocorrelation as well as error bars are calculated by
a bootstrap technique.
We performed 32 randomizations of background galaxy catalogues 
that mimic the noise properties in $\kappa_M$ as predicted by
Eq. \eqref{eq:map-noise-ppties}. We also randomly shuffled the shear
catalogue calculated with Eq. \eqref{eq:lum2shear} before smoothing and
performing the $\gamma$-to-$\kappa$ inversion. Note also that we discarded the
pixels of the convergence maps that lie inside masked areas
(see \reffig{fig:ms0302BVRfield}).
In these regions, the lack of background galaxies severely increases the noise
level. Field boundaries are masked in the same way.\\
In the following, $\mycovv{M}$, $\mycovv{E}$, and $\mycov{M}{E}$ refer to the
mass-mass, light-light  and mass-light correlation functions respectively.
$\mycov{M}{E}$ shows a maximum at zero lag, which is significant at the 10-$\sigma$
confidence level. The cross-correlation peak is fairly isotropic and well
centered on the origin. At zero lag, the normalization parameter of Eq. 
\eqref{eq:general-model} yields $M/L = 277 \pm 27\mslsun$.\\
We thus increased the number of constraints by considering the whole correlation
function profile over the 7 inner arcminutes. The $\lambda$ value
is derived by performing a global $\chi^2$ minimization over the correlation 
functions, using sufficiently sparse sampling points to 
reduce the correlations between bins\footnote{1 arcmin is the
  characteristic length of our spatial smoothing. We checked that the crossed
  terms in the covariance matrix drop significantly beyond this scale.}.
$\lambda$ satisfies the system:
\begin{equation}\label{eq:basic-system}
  \begin{split}
    \mycovv{M}=& \lambda^2 \mycovv{E} \\
    \mycov{M}{E}=& \lambda \mycovv{E}.
  \end{split}
\end{equation}
We found $M/L = 286 \mypm{34}{39} \mslsun$ with
$\chi^2/{\rm dof}\simeq0.88$.
The left panel of \reffig{fig:correl} shows the $\mycov{M}{E}$ and $\mycovv{E}$
correlation profiles with this mass-to-light ratio normalization.
We also observe an excess of light autocorrelation at $\theta \ge 15\arcmin$
which is the characteristic distance between clusters.
Note that this bump is enhanced if we only consider supercluster
early-types and discard the less clustered foreground contribution.

So far, we find an excellent matching between the $\mycovv{M}$, $\mycov{M}{E}$
and $\mycovv{E}$ correlation functions profiles up to $\sim 10$ arcmin.
The linear relation \eqref{eq:general-model} turns out to be a good model.
As already pointed out by K98, the main conclusion is that {\sl early-types
galaxies trace the mass}. Oscillating patterns around the light
autocorrelation appear for $r\ge 8\arcmin$.
 G02 as well as \citet{wilson01} found similar patterns. 
 They are likely  noise artifacts.

As compared to the results of \refsec{subsec:clust_proper}, the correlation
analysis gives a value for the mass-to-light ratio $\simeq$ 15\% higher than
that deduced from integrated quantities inside one megaparsec around clusters.
The $M/L$ deduced from $\kappa$ maps is insensitive to a constant mass sheet
(the so-called mass-sheet degeneracy).
Therefore, it is necessary to subtract the mean luminosity contribution in the
circular aperture of individual clusters analysis and to only consider
the excess of luminosity. We find that within 3 arcmin from the center
$M/L = 273 \pm 47 \mslsun$. Therefore, the agreement with the overall correlation
analysis is excellent.  

The agreement with the K98 results, after rescaling to a flat $\Lambda$
cosmology, is also excellent. Their conclusion that early-type galaxies
trace the mass faithfully is therefore confirmed by our analysis.
Nevertheless, the authors argued that they saw little evidence for any variation
of $M/L$ or 'bias' with scale. K98 addressed this issue by performing the 
correlation analysis in the Fourier space by splitting the data into a low
and a high frequency bin. They found an increase of $M/L$ ratio with increasing
wavelength, ranging from $\sim 180$ at scales $\lesssim 2.5\hmMpc$ to
$\sim 280$ beyond. The physical meaning of this trend is not clear.
Variations of $M/L$ ratio with scale likely indicate underlying physical changes
in the relations between mass and light that cannot be interpreted from
our simple linear scale-free biasing parameter $\lambda$.  
In the following, we investigate some models that may explain the 
$M/L$ variations observed by K98.

\subsection{Changing the dark matter halo profile}\label{subsec:haloprofile}
In Eq. \eqref{eq:shear-ptmass}, we assumed that dark matter halos of
individual galaxies have a little extension compared to the weak
lensing filtering scale, so that they can be modeled as point masses with
mass proportional to the galaxy luminosity. In this   
section we study how a more complex dark matter halo profile may 
change the conclusions of the previous section.\\
Let us consider a truncated isothermal sphere (TIS)
\citep{brainerd96,schneider97}. The convergence reads
\begin{equation}\label{eq:trunc-conv}
  \kappa_{\rm TIS}(r) = \frac{b}{2 r} \left[ 1 - \frac{r}{\sqrt{r^2+s^2}}\right].
\end{equation}
where $s$ is the truncation radius. When $s\rightarrow \infty$, $b$ reduces to
the Einstein radius $\theta_E$ of the singular isothermal sphere (SIS).
Assuming a $L \propto \sigma^4$ scaling relation \citep{faber76,fukugita91}
and $M_{\rm tot}=\pi \scrit b s \propto L$, we set
\begin{equation}\label{eq:scaling-param}
  \frac{b}{b_*}=\left(\frac{\sigma}{\sigma_*}\right)^2=\left(\frac{L}{L_*}\right)^{1/2}\;,\quad \frac{s}{s_*}= \left(\frac{L}{L_*}\right)^{1/2}\;.
\end{equation}
This empirical parameterization is consistent with  \citet{wilson01a} who
assumed $b\propto L^{1/2}$, as well as with \citet{hoekstra03} who found
$b\propto L^{0.60\pm0.11}$ and $s\propto L^{0.24\mypm{0.26}{0.22}}$ leading to
$M\propto L^{0.84\mypm{0.28}{0.25}}$.
\begin{figure*}[htb]
  \includegraphics[width=6.2cm]{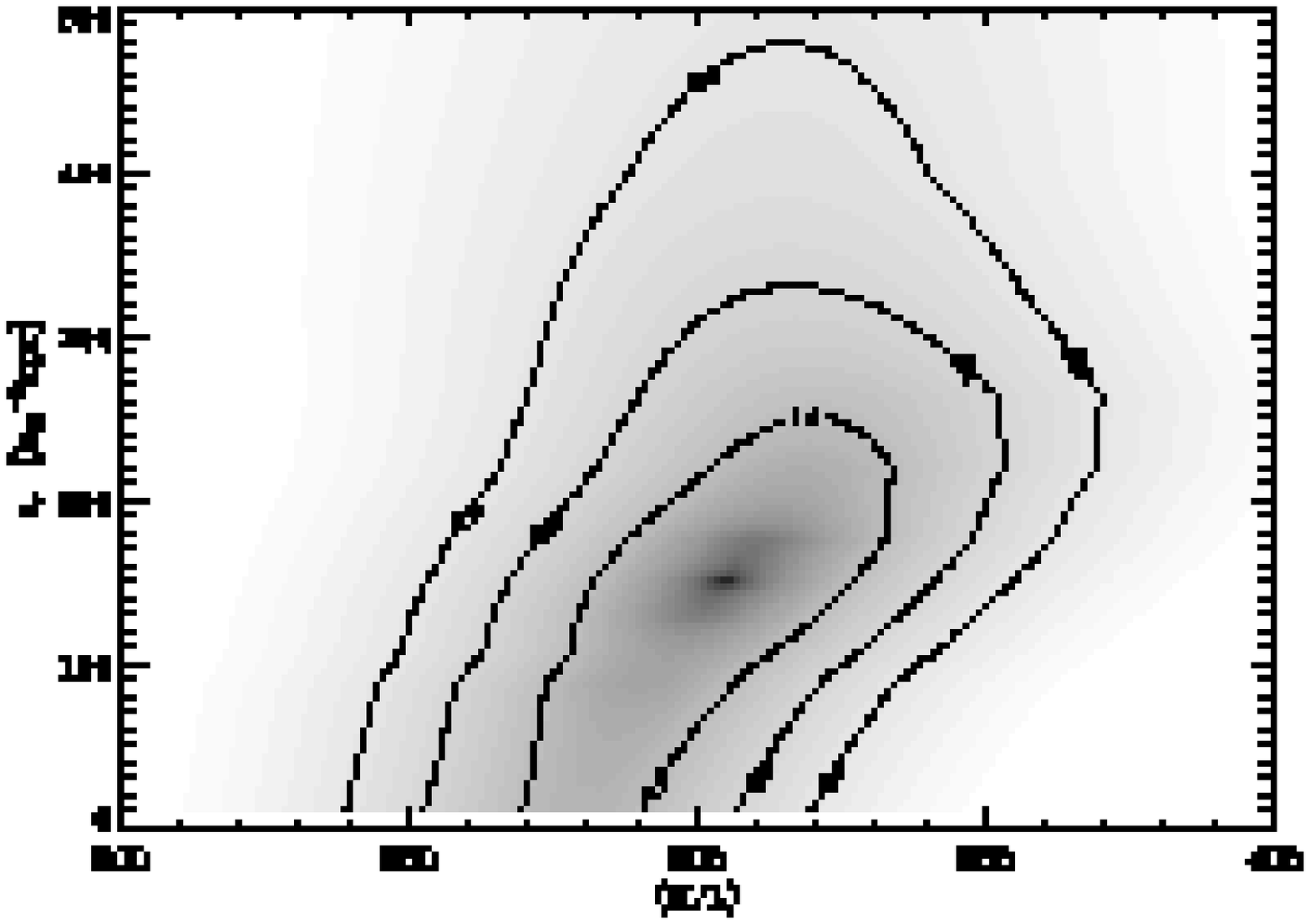}
  \includegraphics[width=6.2cm]{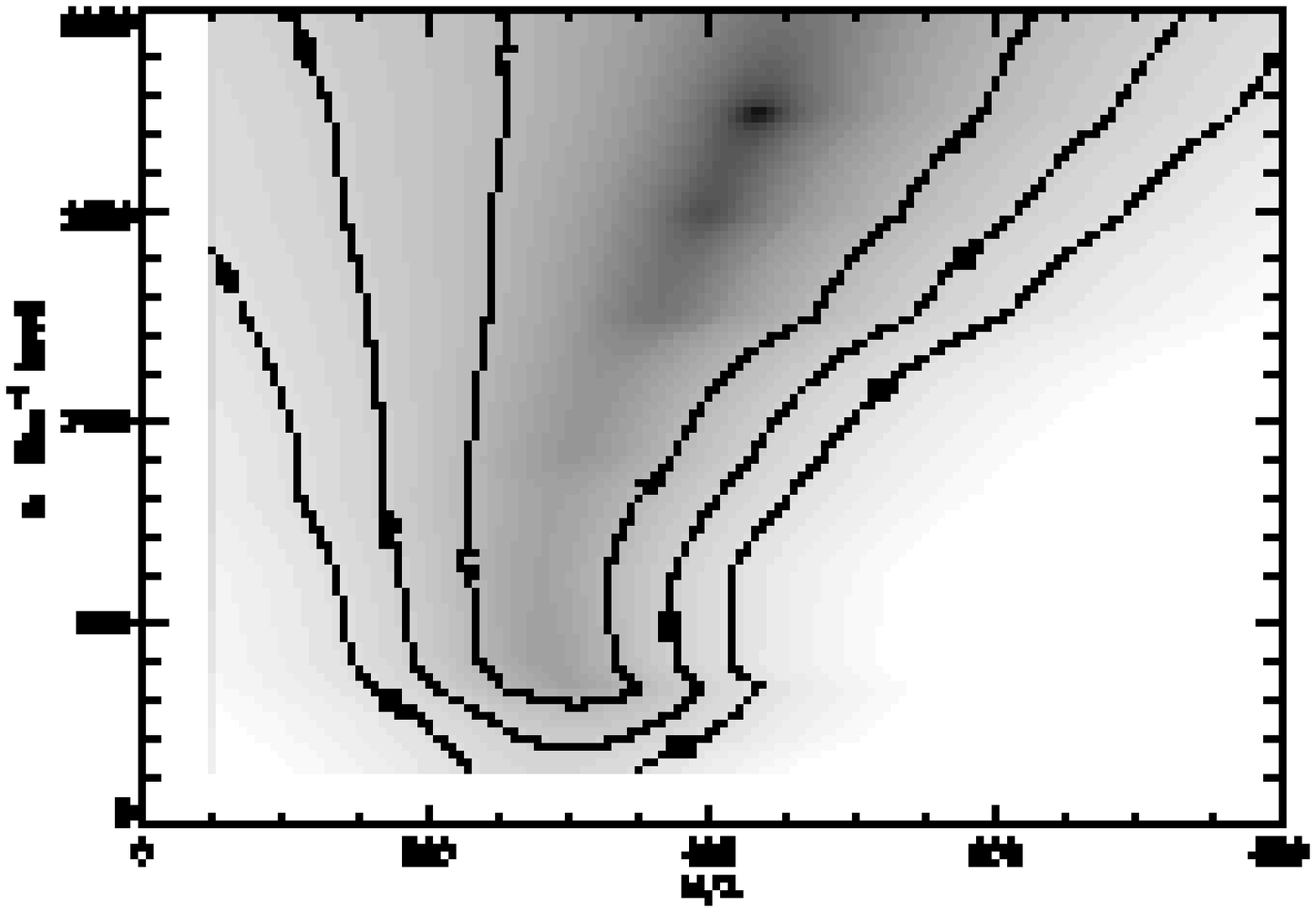}
  \includegraphics[width=6.2cm]{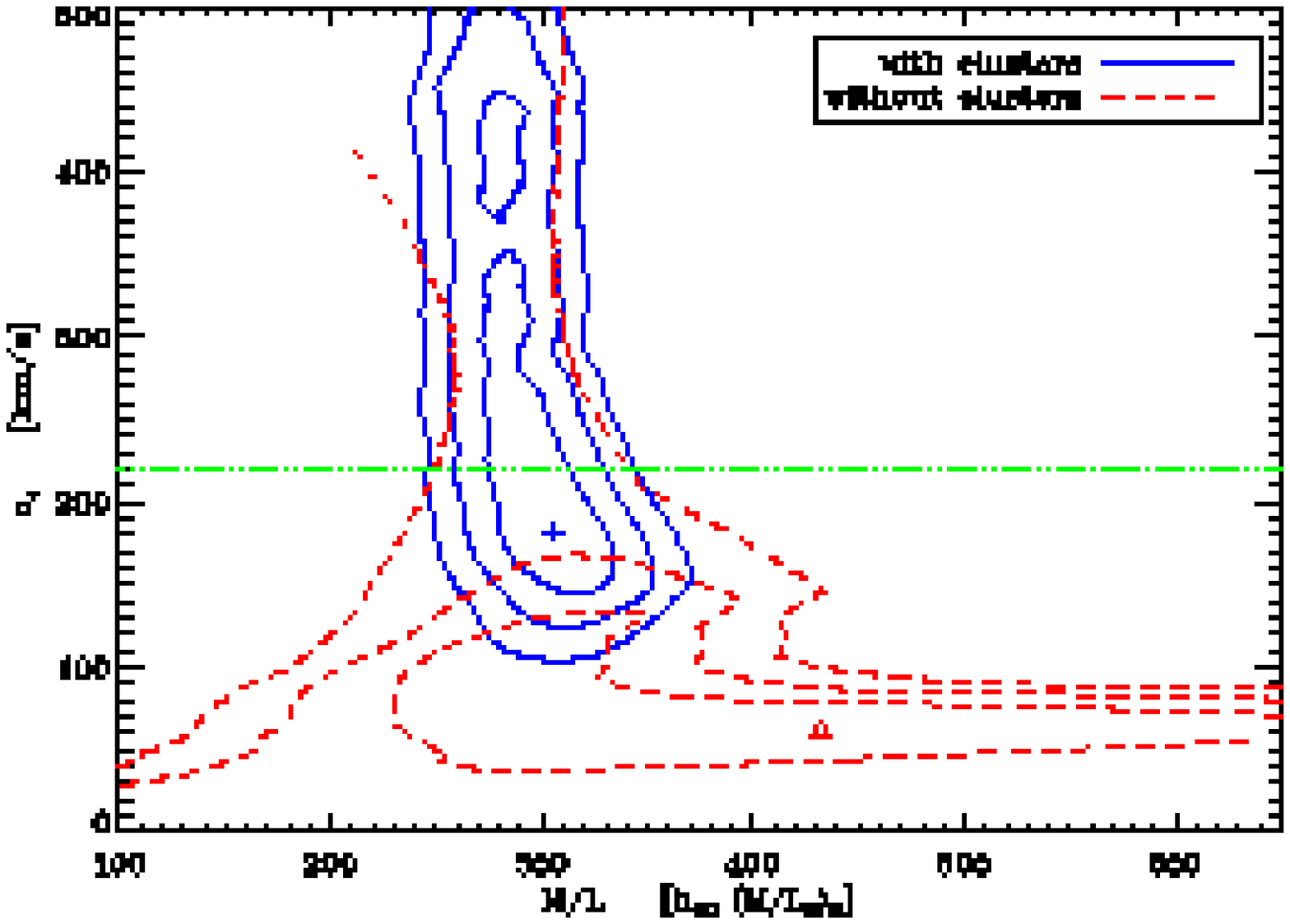}
  \caption{TIS halo modeling.
    Left : Contour plot showing constraints
    of mass-to-light ratio and truncation radius of a $L_*$ early-type galaxy.
    The data are consistent with dark matter halos with truncation radius
    $s_*\lesssim 200\hmkpc$.
    Middle : same plot when considering the periphery of clusters only,
    the tendency is reversed and large halos ($s_*\gtrsim 400\hmkpc$) are favored.
    Right : same constraints interpreted in the $\sigma_*,M/L$ plane.
    The two cases with/without cluster masking are overlaied (solid blue / 
    dashed red contours respectively). The horizontal straight line
    $\sigma_*=220 \,{\rm km\,s^{-1}}$ is a  fiducial value for local elliptical
    galaxies.}
  \label{fig:contour_trunc}
\end{figure*}
Given that $b_* = (M/L) \frac{L_*}{\pi s_* \Sigma_{\rm crit}}$,
we have to constrain the pair $(M/L,s_*)$, or equivalently
$(\lambda= (M/L)/300,s_*)$. $\lambda$ no longer contributes linearly to the
$\kappa_E$ expression because of the dependence of $s$ on $L$.\\
The correlation functions are calculated in the same way as in
\refsec{subsec:lin-bias-cor}. However, since $s$ is different from one galaxy to
another, the resulting correlation function $\mycov{M}{E}_{\rm TIS}$
(resp. $\mycovv{E}_{\rm TIS}$) is no longer the convolution of $\mycov{M}{E}$ 
(resp. $\mycovv{E}$) by the normalized halo profile (resp. normalized halo profile
autocorrelation), making the CPU cost much more important.

Contour plots for $(M/L,s_*)$ are displayed in the left panel of
\reffig{fig:contour_trunc} yielding $(M/L)_{\rm TIS} = 305 \mypm{30}{35} \mslsun$
and $s_* = 150\mypm{90}{150}\hmkpc$. 
This value is smaller but still statistically consistent with
$s_*=264\pm{42} \hmkpc$ derived by \citet{hoekstra03}. 
However, \citet{hoekstra03}  used both early and late type field galaxies 
and also relaxed the constraint $M \propto L$, making a comparison 
with our sample difficult. However, because we are using a smoothing scale
$\theta_s=40\arcsec=220\hmkpc$, it is only possible to put an upper limit
$s_* \lesssim 200 \hmkpc$. We therefore cannot rule out that 
tidal stripping effects in dense environments may decrease the galaxy cut-off 
radius, as reported by \citet{natarajan02}. This point will be discussed
in more detail in the next sub-section.

It is also interesting to interpret our results in terms of halo velocity
dispersion $\sigma_*=c\sqrt{\frac{b_*}{4 \pi D_d}}$ as shown
in the right panel of \reffig{fig:contour_trunc}. The Results are consistent
with general values for $\sigma_*$
\citep[see \eg][and references therein]{seljak02}.

\subsection{Large scales / Periphery of clusters}\label{subsec:outer-cor}
The results derived in \refsec{subsec:lin-bias-cor} and
\refsec{subsec:haloprofile} are in good agreement with those of
\refsec{subsec:clust_proper}. They confirm that the average  
mass-to-light ratio of halos is $M/L \approx 300\mslsun$ and that early type
galaxies are the primary tracers of dark matter on supercluster scales.
Their contribution may however depend also on the local density,
and the average value we derived could only reflect a biased signature
of the mass-to-light ratio dominated by the three clusters. One could
conclude equally well either that early-types trace the mass at all scales
with a constant $M/L=300\mslsun$ or that the signal coming from clusters
is too strong, and hides more subtle details. This would explain why K98
reported an increasing variation of $M/L$ ratio with increasing scale
using two bins of low and high spatial frequencies.

To clarify this, we calculate the correlation functions as above,
but we discard the central regions of clusters. More precisely,
we set to zero the inner 3 arcmin around each cluster
(circles of \reffig{fig:ms0302BVRfield}) to compute the residual
correlation produced by the larger scale structures, like filaments and voids.
When considering the periphery of clusters only, the amplitude of correlation
functions drops by a factor $\sim 3$ showing that most of the signal comes
from the clusters.

\begin{figure*}
  \includegraphics[width=18cm]{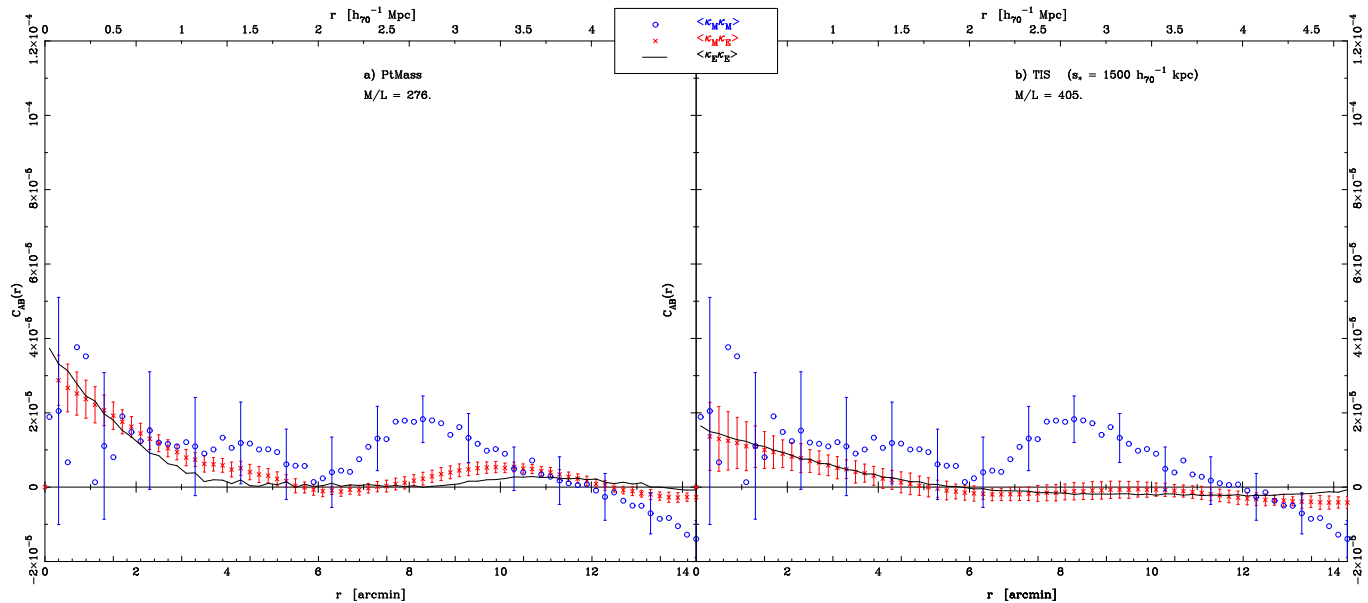}
  \caption{Same as \reffig{fig:correl} but with the clusters centers. A large
    truncation radius TIS model (right panel) provides a better fit than the
    point-mass halo model (left panel) which does not fit the data well
    ($\chi^2/{\rm dof} \simeq 2.1$). Notice how compared to \reffig{fig:correl},
    the correlation amplitudes drop when the clusters centers are discarded.}
  \label{fig:correl-trous}
\end{figure*}

The constant $M/L$ ratio model with point-mass-like dark matter halo 
provides a rather bad fit:  $\chi^2/{\rm dof} \simeq 2.1$. This is
significantly worse than for the whole field analysis. The best fit yields
$M/L = 276 \pm 30\mslsun$ and is plotted in the left panel of
\reffig{fig:correl-trous}. For a TIS halo model the goodness-of-fit is significantly
improved $(\chi^2/{\rm dof} \simeq 0.65)$ when constraining $(M/L,s_*)$.
However, it  requires $s_* \gtrsim 300 \hmkpc$ and
$M/L = 280 \pm 40$. Note that large values of $s_* \sim 1.5\hmMpc$ with larger
$M/L \sim 400$ are also consistent with the data (see middle panel of
\reffig{fig:contour_trunc}). The right panel of \reffig{fig:correl-trous}
shows such an extended halo profile. It is worth noting that these solutions
appear unphysical and may rather indicate that the input model is not well suited. 

The fact that halos are more extended outside the cores of clusters
is also consistent with the tidal stripping hypothesis. 
However, as discussed in the following section, this conclusion depends
on the input model $M \propto L$ and on the fact that we assumed that all
the mass is associated with early-type galaxies. In particular, the
contribution of late-type galaxies has been neglected again.
The small amount of mass located in low frequency modes in the the outer
parts of the supercluster may give an indication that
these modes are not well traced by early-type galaxies.

\section{Discussion}\label{sec:summary}
The MS0302+17 supercluster mass distribution, derived from weak
lensing analysis of background galaxies, matches
the supercluster light distribution of its early type galaxies.
The correlation between them is very strong. More precisely
the shape of the light-mass cross-correlation profile is in excellent
agreement with a simple model where dark matter is directly related
to light, assuming a constant mass-to-light ratio.\\
We  confirm the results of K98, with different data sets and a larger field, 
with different hypotheses to derive lenses samples and as well as background
sources catalogues, and by using an independent PSF correction method.
Therefore, the strong correlation they found
is confirmed and strengthened by this work. In particular
we confirm that $M/L \simeq 300 \pm 40\mslsun$ with most matter attached
to the early-type galaxies. A generalization of our findings to all
supercluster systems is premature but it is worth mentioning that
\citet{wilson01} and G02 found similar trends in blank fields and the
A901/A902 supercluster, respectively.

When we introduce dark matter halos in the form of truncated isothermal
spheres (TIS) we show that the linear relation $M \propto L$ is still
verified and that dark matter halos of early-type galaxies must be rather compact
($s_* \lesssim 200\hmkpc$) near the cluster centers that dominate the signal.
We attempted to mask the clusters centers to analyse the remaining signal.
Removing these regions before doing the correlation analysis gives indications
that galaxy halos are more extended at the periphery of clusters than in the inner
regions. Such a behavior is also consistent with the previous result of K98
who measured a different $M/L$ ratio when considering low and high spatial
frequency modes of $\kappa$ and $\kappa$-from-light maps. This result,
which does not have a straightforward physical explanation, together with
our halo analysis, can be interpreted in two different ways provided
$M/L$ is constant and dark matter halos follow a nearly TIS density profile:
\begin{enumerate}
\item either most of the mass is attached to early-type galaxies with 
  $M/L=300\mslsun$ and is distributed into halos that are more compact
  when located closer to clusters cores
  (consistent with the tidal stripping hypothesis);
\item or, at the periphery of clusters $M/L={\rm constant}$ is not completely
  verified. Late type galaxies (which are more abundant at the periphery of
  clusters) or a more diffuse dark matter component that does not follow the
  light from early-types in a simple manner may likely be an increasingly 
  important mass component beyond the cluster scale.
\end{enumerate}

\paragraph{Do late type galaxies contribute to the supercluster mass?}
The relation between dark matter and light distribution of late type galaxies
is difficult to derive from our data only. Late type supercluster galaxies
cannot be easily extracted from our galaxy color-color diagrams, because
the B and V data are not as deep as the R image. Furthermore, the color-color
tracks of late type galaxies are broader than those of early type galaxies and
are therefore much more difficult to separate from field galaxies. Nevertheless,
we find that the relation between dark matter and late type galaxies is weaker
than and possibly different from that of the sample of early-type galaxies.
The cross-correlation profile can be interpreted as if only a small amount of
matter is associated to these galaxies. A contribution of low frequency
modes to the correlation functions is not surprising since there is
compelling evidence that late-types are much less clustered
and less massive than early-types \citep{Budavari03}.\\
From a lensing analysis point of view, it is therefore expected that the
convergence $\kappa_{\rm late-types}$ is localized in low frequency modes,
on scales that could be similar to the CFH12K angular size.
A weak lensing analysis on a single field may not be relevant for probing the
mass-light cross correlation of late type galaxies on the supercluster scale.
A much larger field of view is likely needed. \citet{Gray03}
used COMBO-17 data in the Abell901/902 system and found that late-types
are basically located in underdense regions, a result compatible with the
well known segregation effect.

On cluster scales the three systems ClN, ClS and ClE
show clear lensing features (arcs or arclets). Their properties show
a scatter of mass-to-light ratios with various morphological aspects
(with indications that ClS and ClN seem dynamically relaxed and ClE has
ongoing merging-like events). Overall, the supercluster
dynamical state may indicate that ClE could not be gravitationally bound
to the whole system. A large number of redshifts in the field would be useful
to confirm the global dynamical stage of this system.\\
K98 possibly detected a filament of dark matter connecting ClS and ClN.
We do not confirm this. We just observe an elongated
structure in the $\kappa$-maps which is located westward of ClS and is
likely due to a casual projection effect creating a bridge between ClS
and a clump probably belonging to the supercluster. Another filamentary
structure extends toward the West of ClN along the field boundary.
We indeed also observe a visible counterpart in the $\kappa_E$ maps.
Finally, there is no conclusive evidence for any detection of
filamentary structure in the field of MS0302+17.
The detection of K98 may be due to residual systematics in the PSF
anisotropy correction process. Note also that G02 observed a filament
connecting A901b and A901a but it was not confirmed by an optical counterpart.
Indeed, a detection of dark matter dominated filaments similar to what is
seen in numerical simulations remains challenging for such lensing studies.

Finally, we also observe a large under-dense region located between the three
clusters. Its angular size is about $\sim 12\arcmin$. The depression
amplitude is $\Delta \Sigma \approx -3\times10^{12}\hmMsun/{\rm arcmin}^2$.

There is an observational issue that needs clarifications. 
In A901/A902, G02 derived $M/L \sim 88\mslsun$ with correlation analysis
whereas they found $M/L \sim 140 \mslsun$ in apertures around clusters.
\citet{wilson01} derived a constant $M/L = 210 \pm 53 \mslsun$ for 
their blank fields sample. These values are significantly different
from K98 and this work. The reason for this discrepancy is not clear. 
The large scatter in $M/L$ found by G02 from one cluster to another 
may be intrinsic if one assumes that each cluster is in a  
different dynamical stage. G02 investigated whether the large
scatter could be interpreted as a natural scatter in the
mass/light relation. Using the \citet{Dekel99} formalism, 
they claimed they measured a marginal nonzero stochastic component in
the Abell901/902 system. In the case of MS0302+17, we are unable to measure
such a positive stochastic term in the correlation function profiles.
The subcomponents of the A901/902 supercluster are physically closer than those
found in MS0302+17. The average projected physical separation of the former
is of order $2\hmMpc$ whereas the MS0302+17 clusters are separated
by $\sim 5\hmMpc$ showing that possible interactions and dynamical
stages are different from one supercluster to another.

\section{Conclusions}\label{sec:conclusion}
We have analyzed the weak lensing signal caused by the supercluster of
galaxies MS0302+17 and connected it to its optical properties.
Using a BVR photometric dataset from CFH12K images, we 
identified the early type members of the supercluster. The R band image was also
used to measure the coherent gravitational shear produced by massive structures
of the supercluster and by foreground contaminating field objects.\\
When considered individually, each cluster has an average rest frame B band
mass-to-light ratio $M/L=249 \mypm{41}{32}\mslsun$. The Eastern cluster does
not show a well relaxed structure. It can be viewed more likely as a two 
component cluster system with ongoing gravitational interaction. This may
explain the rather poor agreement between lensing and kinematic estimates of the
velocity dispersion. It also supports the previous conclusions of
\citet{fabricant} and K98 that ClE may not be gravitationally bound to
the system made of the other two ClN and ClS clusters.\\
The mass (or convergence) map shows an excellent agreement with that derived from
the distribution of early-type galaxies. Besides the well detected main clusters,
one can observe a large underdense region between them. We were unable to
confirm the existence of a filament joining ClS and ClN as claimed by K98.\\
We performed a correlation analysis between ``light'' and mass aiming at
probing whether the linear relation $M \propto L$ (or more precisely
$\kappa \propto \kappa_{\rm from\ light}$) is consistent with the data at hand.
The results based on mass-mass, mass-light and light-light correlation functions
are robust enough to make conclusive statements on the average mass-to-light
ratio. We found that $M/L\simeq300\mslsun$. In other words, all the mass
detected from weak lensing analysis is faithfully traced by the luminosity
distribution of early-type galaxies. 

Our conclusions are in excellent agreement with those of K98.
They only depend slightly on the unknown distribution of late-type galaxies,
since their contribution is found to be small. However, when focusing on
early types, we were able to put constraints on the density profile of
galaxy halos. Despite the rather large spatial smoothing, we conclude that halo
truncation radii $s_* \lesssim 200 \hmkpc$ for an $L_*$ galaxy.
We also found evidence for a relaxation of this constraint at the periphery of 
clusters: $s_* \gtrsim 300 \hmkpc$. However, this latter result relies on the
fact that late-type galaxies are neglected. Such an hypothesis may not be so
evident at large distance from the centers of clusters.

Further investigations of the MS0302+17 supercluster of galaxies may
require more photometry (in different optical/NIR bands) in order to identify
late-type galaxies in the supercluster and compare their distribution and
physical properties with the early-type sample.
 
\begin{acknowledgements}
  We thank E. Bertin,  H. J. McCracken, D. Clowe, N. Kaiser and
  L. van Waerbeke for useful discussions, and D. Leborgne for providing
  galaxy evolution tracks and K-corrections. We also thank T. Hamana for
  fruitful comments and a careful reading of this paper.
  We are also thankful to the anonymous referee for useful comments.
  The processing of CFH12K images was carried out at the TERAPIX data
  center, at the Institut d'Astrophysique de Paris. Y.M. and
  some processing tools used in this work were partly funded by
  the European RTD contract HPRI-CT-2001-50029 "AstroWise".
\end{acknowledgements}

\bibliographystyle{aa}
\begin{scriptsize}
  
\end{scriptsize}

\end{document}